\documentclass[aps,pra,twocolumn,amsmath,amssymb,floatfix]{revtex4-2}

\usepackage{graphicx}
\usepackage{dcolumn}
\usepackage{bm}
\usepackage{psfig}

\begin{document}

\newcommand{\ea}{{\it et al.}}

\preprint{APS/123-QED}

\title{Carbon elastic and inelastic stopping-power components \\
 for heavy ions at Bohr and higher velocities}

\author{R.N. Sagaidak}
 \email{sagaidak@jinr.ru}
 \affiliation{Flerov Laboratory of Nuclear Reactions, Joint Institute for Nuclear Research, J.-Curie, 6, 141980 Dubna, Moscow region, Russia}

\date{\today}

\begin{abstract}
Carbon stopping-power (SP) data for heavy ions (HIs), obtained around Bohr velocities, revealed remarkably lower values than those predicted by the SRIM/TRIM calculations/simulations. An attempt was made to extract the elastic (collisional) and inelastic (electronic) components from available SP data. A problem is that essentially total SP is  measured in experiments, whereas electronic SP values, presented as the results, are obtained by the subtraction of the calculated collisional component values from the measured ones. At high HI reduced velocities $(V/v_{0})/Z^{2/3}_{\rm HI} \gtrsim 0.3$ ($V$ and $v_{0}$ are HI and Bohr velocities, respectively, and $Z_{\rm HI}$ is the HI atomic number), the collisional component can be neglected, whereas at Bohr velocities it becomes comparable to the electronic one. These circumstances were used to compare  the experimental data with SRIM/TRIM calculations/simulations and to extract empirically the collisional SP component.
\end{abstract}
\maketitle

\section{\label{Intromotiv}Introduction}

The stopping power (SP) of solids for heavy ions (HIs) passing through thin foils is an important value that may serve as a basis for the description of specific ion-atom interactions. Large amounts of SP data accrued-to-date by James F. Ziegler \cite{SRIM} and Helmut Paul \cite{IAEASP} allow us to compare the data with different calculations developed to reproduce the experimental data for practical usage. The statistical analysis of their applicability for HI stopping in elemental solids, performed in \cite{IAEASP,Paul2003,Paul2010,Paul2013}, shows that mean normalized deviations for SP values at the lowest range of HI energy $E$ = 0.001--0.25 MeV/nucleon are the smallest for the SRIM code \cite{SRIM} compared to other approaches.

It should, however, be noted that the considerations \cite{Paul2003,Paul2010,Paul2013,IAEASP} dealt with the electronic component of SP ($SP_{e}$) in contrast to the total SP value ($SP_{tot}$) measured in experiments. The latter is usually considered as the sum of electronic and nuclear $SP_{n}$ (collisional) components while analyzing the data. SRIM calculations give us both the components and show that the contribution of $SP_{n}$ to $SP_{tot}$ becomes significant at $E \lesssim 0.1$ MeV/nucleon. At the same time, the qualitative analysis \cite{Paul13AIP} has shown, that the nuclear component calculated with SRIM $SP_{n}^{\rm SRIM}$ may overestimate its contribution to the $SP_{tot}$ values, as obtained for some media and HIs in the energy range of  $0.01 \lesssim E \lesssim 0.1$ MeV/nucleon \cite{Zhang2002,Barb2010}.

According to the databases \cite{SRIM,IAEASP}, a carbon SP data set is the largest one. In the early experiments on carbon SP measurements for HIs at Bohr \cite{Fastrup66,Hvelp68,Lennard86} and higher velocities \cite{BrownMoak72}, the $SP_{e}$ values were determined by subtraction of  $SP_{n}$ from $SP_{tot}$, as mentioned above. The $SP_{n}$ values were calculated \cite{Fastrup66,Hvelp68} or obtained with Monte Carlo (MC) simulations \cite{Lennard86}. The measurements were carried out for HIs escaping  targets within narrow angle $\theta_{\rm out}$ relative to the beam direction crossing a target ($\theta_{\rm out}\leqslant 1/3^{\circ}$ \cite{Fastrup66,Hvelp68} and $\theta_{\rm out}\leqslant 0.17^{\circ}$ \cite{Lennard86}). The estimates of $SP_{n}$ contributions showed remarkable values at the lowest energies \cite{Fastrup66,Hvelp68} and for relatively thick targets \cite{Lennard86}. An intercomparison of the HI energy losses caused by the elastic collisions, as estimated with the calculations \cite{Fastrup66} and MC simulations \cite{Lennard86,Krist84}, showed their differences from each other within a factor of 1.5--3. The $SP_{n}$ dependencies on the HI detection aperture followed from the analysis of the experimental data \cite{Garnir80} and the theoretical work \cite{Krist84}.

This work attempts to re-estimate the contribution of nuclear stopping to the total carbon SP for HIs at Bohr and higher velocities. A direct comparison of SP data obtained at HI velocity $V = 0.8 v_{0}$ with the results of SRIM/TRIM calculations/simulations is considered in the next section. In Section~\ref{totSP}, total and electronic SP values as a function of the HI velocity will be compared with SRIM calculations, and nuclear SP values will be estimated. Nuclear SP estimates thus obtained will be discussed and some conclusions will be made in Section~\ref{discus}.

\section{Stopping power data survey at $0.8 v_{0}$}

The available SP data at low HI velocities \cite{Fastrup66,Hvelp68,Lennard86} provide a good opportunity to check their reproducibility within different approaches. The data of Lennard \ea\ \cite{Lennard86} for HIs from F to U at the velocity of $0.8 v_{0}$ were obtained using two sets of carbon targets (thicknesses $\Delta X \sim$5 and $\sim$30 $\mu$g/cm$^{2}$). The velocity corresponded to the energies $E = E_{in} - \Delta E / 2$ ($E_{in}$ and $\Delta E$ are the input energy and the energy absorbed by the $\Delta X$-thick target, respectively). These values are listed in the respective tables \cite{Lennard86}. In earlier data \cite{Fastrup66,Hvelp68}, SPs were measured for HIs from C to Y in the energy range of 0.1--1.5 MeV, that was around the $0.8 v_{0}$ value. The measurements were performed with targets of different thickness (3.8--24.1 $\mu$g/cm$^{2}$). In our consideration of data \cite{Fastrup66,Hvelp68}, the SP values corresponding to $0.8 v_{0}$ were obtained by the interpolation of the tabulated $SP_{e}$, $SP_{n}$, and $\Delta X$ values.

\subsection{\label{SRIMTRIMcomp}Comparison with SRIM/TRIM calculations/simulations}

\begin{figure}[!h]  
\begin{center}
\includegraphics[width=0.485\textwidth]{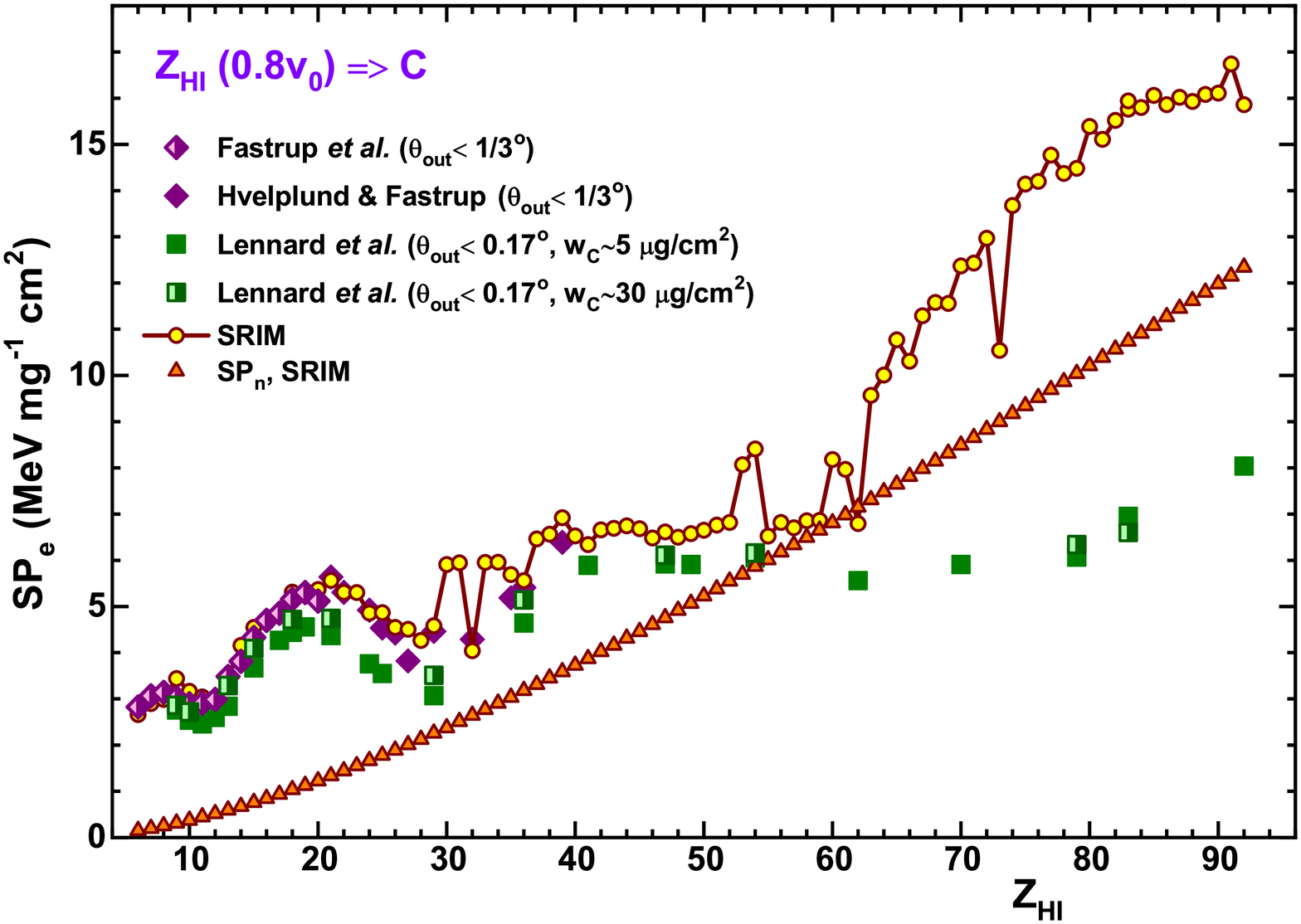}
\caption{\label{SPeSRIMcomp}The electronic SP values ($SP_{e}$) obtained in the experiments \cite{Fastrup66,Hvelp68,Lennard86} (diamonds and squares) and those calculated by SRIM (open circles connected by a solid line) are shown as a function of $Z_{\rm HI}$. The nuclear SP values ($SP_{n}$) calculated by SRIM are also shown by open triangles for the reference.}
\end{center}
\end{figure}

Figure~\ref{SPeSRIMcomp} shows the electronic SP values obtained in the experiments \cite{Fastrup66,Hvelp68,Lennard86} and those calculated by SRIM. As one can see in the figure, SRIM calculations significantly exceed the experimental data for Xe and heavier ions. The excess corresponds to a factor of $\sim$2 for U ions. A possible contribution of the SRIM nuclear stopping is also shown in the figure, which is comparable with the SRIM electronic stopping for mid-heavy ions (30 $\leqslant Z_{\rm HI} \leqslant$ 62) and is a little less than the respective values for heavier ions. The reason for the inconsistencies between the experimental and calculated data could be twofold: the overestimates of the $SP_{n}$ values subtracted from experimental $SP_{tot}$ \cite{Lennard86} or/and similar overestimates in the $SP_{e}$ values calculated by SRIM.

\begin{figure}[!h]  
\begin{center}
\includegraphics[width=0.485\textwidth]{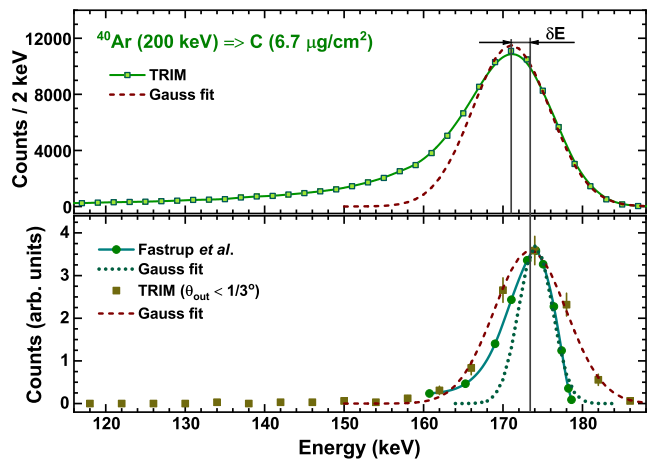}
\caption{\label{ArTRIM}The energy distributions for Ar ions passed through the carbon foil, as obtained in TRIM simulations and in the experiment \cite{Fastrup66}. The upper panel shows the energy distribution for all events, as obtained in simulation (open squares connected by a solid line) and the Gaussian fit to the high-energy part of the distribution (dashed line). The bottom panel shows the energy distribution corresponding to the ions escaping the carbon foil at $\theta_{\rm out} \leqslant 1/3^{\circ}$ (closed squares) and the Gaussian fit to this distribution (dashed line). These are compared to similar ones obtained in the experiment \cite{Fastrup66} (closed circles connected by a solid line) and fitted with the respective Gaussian (dotted line).}
\end{center}
\end{figure}

\begin{figure}[!h]  
\begin{center}
\includegraphics[width=0.485\textwidth]{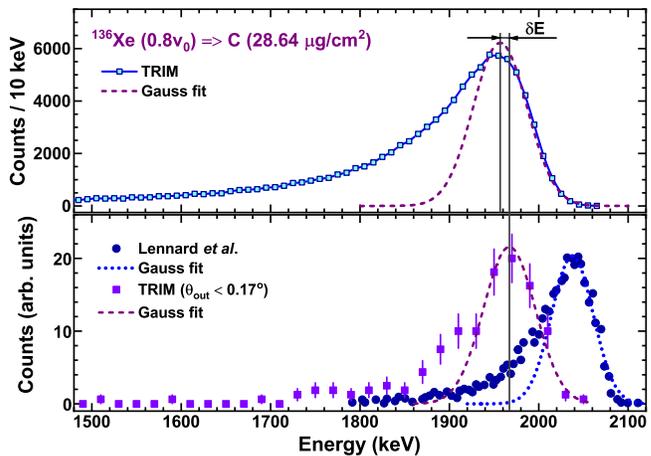}
\caption{\label{XeTRIM}The same as in Fig.~\ref{ArTRIM} but for Xe ions and the experimental data obtained at $\theta_{\rm out} \leqslant 0.17^{\circ}$ \cite{Lennard86}.}
\end{center}
\end{figure}

In an attempt to test these assumptions, TRIM simulations were performed, which allowed us to extract the events corresponding to the definite range of angles for HI escaping stopping foils, and thus to simulate the conditions of the experiments \cite{Fastrup66,Hvelp68,Lennard86}. In Figs.~\ref{ArTRIM} and \ref{XeTRIM}, the results of such simulations are shown for the energy distributions of Ar and Xe passed through the carbon foils of the respective thicknesses. They are compared with the energy distributions obtained for these ions in the experiments \cite{Fastrup66,Lennard86}. The total energy distributions, corresponding to 10$^{5}$ simulations for the ions passing through the foils, and the energy distributions for HIs escaping the foils within output angle $\theta_{\rm out}$, are compared with those obtained in the experiments.

As one can see in the figures, simulations show some shifts to higher energies in the maximum positions for the distributions defined by output angles relative to those corresponding to all angles for escaping ions. The values of these shifts are $\delta E = 2.39\pm0.31$ and $10.4\pm3.0$ keV for Ar and Xe distributions, respectively. With these values, $\Delta E / \Delta X$ is reduced by the same amount of 0.36 keV/($\mu$g/cm$^{2}$). Applying these results to possible SP measurements at all angles, one may expect an increase in the total SP values of 8.3 and 4.6\% for Ar and Xe ions, respectively. In the experiments \cite{Fastrup66,Hvelp68,Lennard86}, the $\Delta E$ values were determined in the same way, i.e., as the difference between the maximum positions for the input and output energy distributions. Note that in Ar simulations, the width of the distribution determined by the output angle is larger than the one obtained in the experiment \cite{Fastrup66}, although their maximum positions are close to each other. As for Xe ions, the situation is reversed, i.e., the widths of distributions are close to each other, although maximum positions differ significantly. The reduced value of the output energy obtained in Xe simulations corresponds to a larger absorption energy (stopping power).

\begin{figure}[!h]  
\begin{center}
\includegraphics[width=0.485\textwidth]{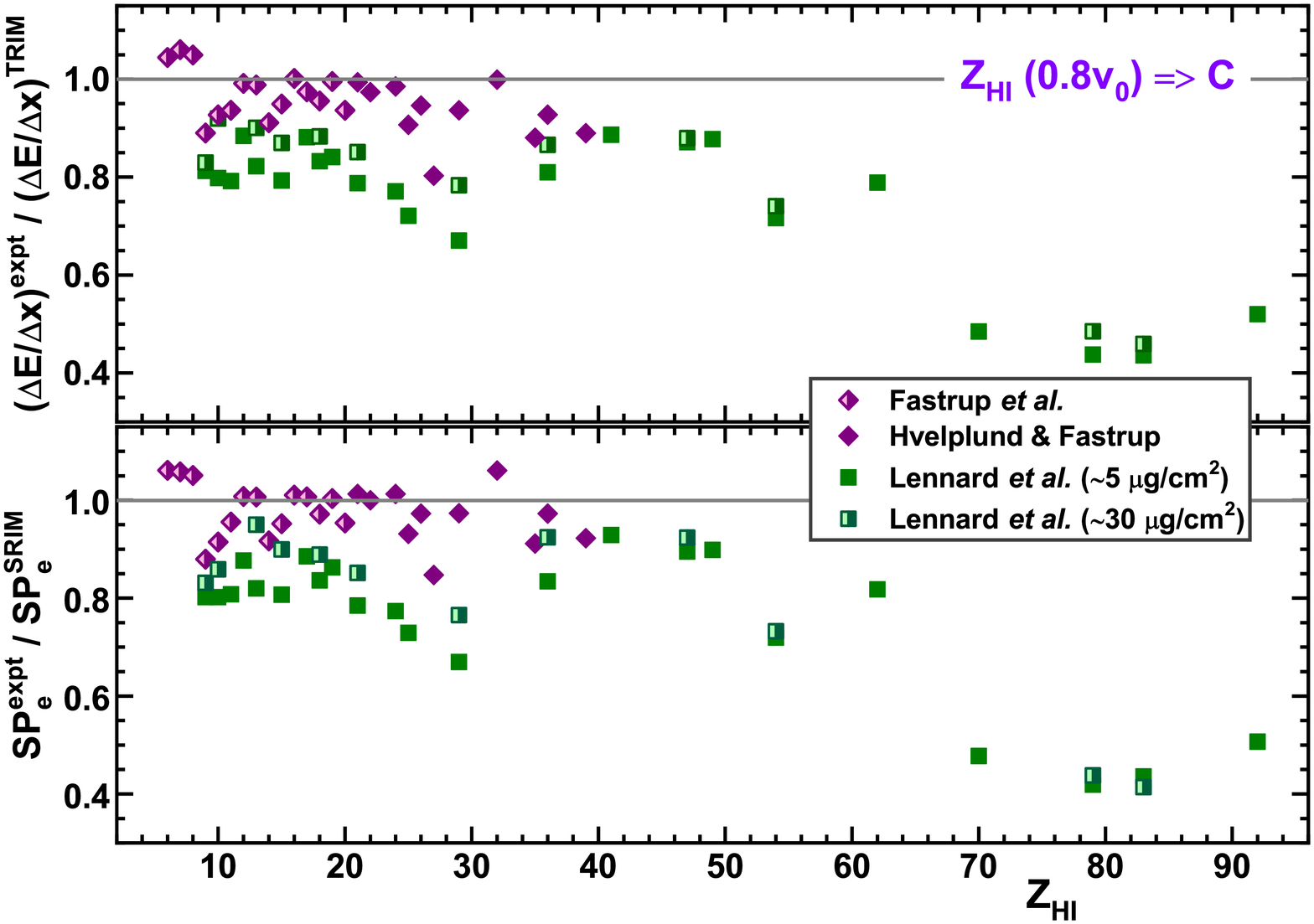}
\caption{\label{dEdXSPecomp}The comparison of the $\Delta E/\Delta X$ values obtained in the experiments \cite{Fastrup66,Hvelp68,Lennard86} to those obtained in TRIM simulations is shown in the upper panel. The comparison of the $SP_{e}$ values extracted in these experiments to those calculated by SRIM (see  Fig.~\ref{SPeSRIMcomp}) is shown in the bottom panel.}
\end{center}
\end{figure}

Figure~\ref{dEdXSPecomp} shows the comparisons of the $\Delta E/\Delta X$ values obtained in the experiments \cite{Fastrup66,Hvelp68,Lennard86} to those obtained in TRIM simulations and of the respective $SP_{e}$ values extracted in these experiments to those calculated by SRIM (both are shown in Fig.~\ref{SPeSRIMcomp}). As one can see, the $(\Delta E/\Delta X)^{\rm expt}/(\Delta E/\Delta X)^{\rm TRIM}$ and $SP_{e}^{\rm expt}/SP_{e}^{\rm SRIM}$ ratios differ slightly. This circumstance means that the angle cutting is insufficient to reduce the impact of nuclear stopping in the calculated/simulated values.

\begin{figure}[!h]  
\begin{center}
\includegraphics[width=0.4\textwidth]{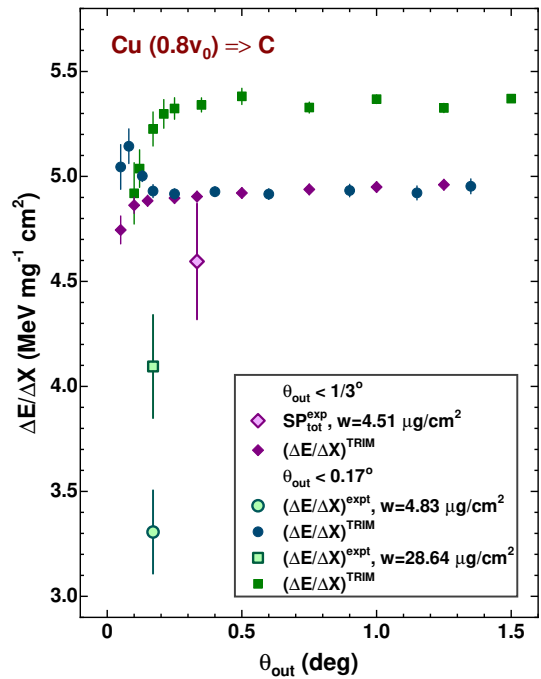}
\vspace*{-10.5mm}
\caption{\label{dEdXTHout}The $\Delta E/\Delta X$ values obtained in the experiments \cite{Hvelp68,Lennard86} (large open symbols) and TRIM simulations (small closed symbols) are shown as a function of output angle $\theta_{\rm out}$ for Cu ions escaping the targets of a different thickness.}
\end{center}
\end{figure}

As was mentioned above, the data \cite{Fastrup66,Hvelp68,Lennard86} were obtained with the targets of varying thicknesses and for varying maximum output angles for HIs escaping the targets. Differences in the experimental conditions of the data \cite{Fastrup66,Hvelp68} and \cite{Lennard86} could be the reason for the data  inconsistencies appeared in Fig.~\ref{dEdXSPecomp}. Figure~\ref{dEdXTHout} shows the $\Delta E/\Delta X$ values obtained with TRIM simulations and in the experiments \cite{Hvelp68,Lennard86} as a function of output angle $\theta_{\rm out}$ for Cu ions escaping the targets of a different thickness. As one can see in the figure, TRIM simulations show the independence of the $\Delta E/\Delta X$ values upon $\theta_{\rm out}$ at $\theta_{\rm out} \gtrsim 0.15^{\circ}$ and $\gtrsim 0.2^{\circ}$ for thin ($\sim$5 $\mu$g/cm$^{2}$) and thick ($\sim$30 $\mu$g/cm$^{2}$) targets, respectively. The experiments with thin targets show that, in going from $\theta_{\rm out} = 0.17^{\circ}$ to 1/3$^{\circ}$, the $\Delta E/\Delta X$ value increases by a factor of $1.39\pm0.12$, whereas the similar value does not change in simulations. At the same time, the ratio of the $\Delta E/\Delta X$ values obtained with thick and thin targets in TRIM simulations ($1.06\pm0.02$) is close to the one obtained in the experiments ($1.24\pm0.11$) \cite{Lennard86}.

Concluding this part, one can state that SRIM/TRIM calculations/simulations essentially overestimate carbon stopping powers obtained for HIs at the velocity of $0.8 v_{0}$ \cite{Lennard86}. At the same time, the electronic SP data for C to Y ions \cite{Fastrup66,Hvelp68} are in satisfactory agreement with SRIM calculations. Further, in attempts to reproduce the data \cite{Fastrup66,Hvelp68,Lennard86}, they are considered within the LSS \cite{LSS63} and other approaches as an alternative to SRIM. Various approximations to the nuclear SP will be also tested.

\subsection{\label{LSScomp}Comparison with LSS and other approaches}

The stopping power considered in the framework of the LSS approach \cite{LSS63} is treated in a similar way as within SRIM, i.e., as the sum of electronic $SP_{e}$ and nuclear $SP_{n}$ components:
\begin{equation}\label{SPtot}
  SP_{tot} = SP_{e} + SP_{n}.
\end{equation}
In the literature, one can find many works aimed at improving each of the components in order to obtain best agreement with updated data supplied by ongoing experiments (see, for example, Ref.~\cite{Dib15} that considers the LSS $SP_{e}$ improvement for polymeric foils).

\begin{figure}[!h]  
\begin{center}
\includegraphics[width=0.485\textwidth]{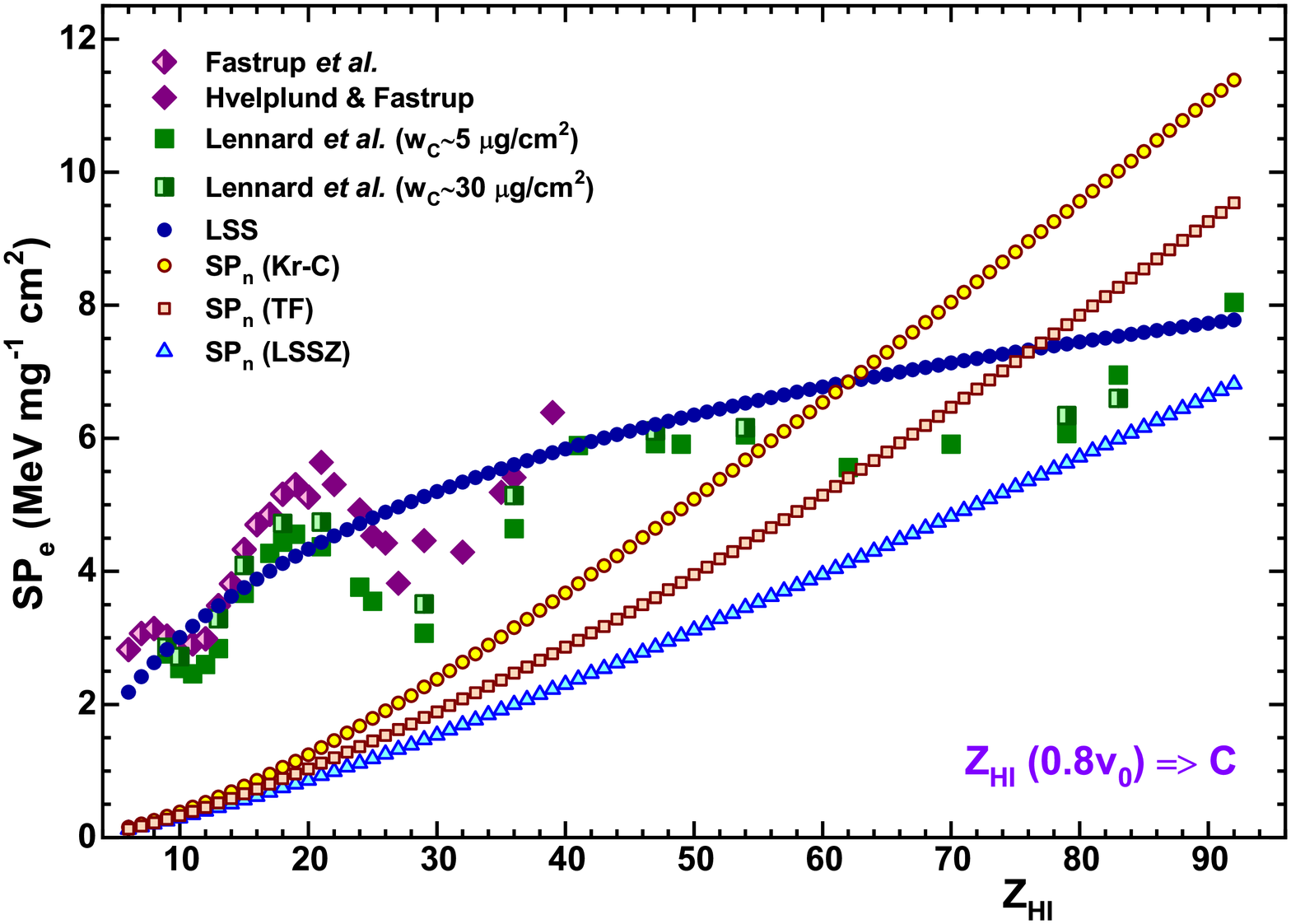}
\caption{\label{SPeLSScomp}The same as in Fig.~\ref{SPeSRIMcomp} but for $SP_{e}$ comparison to the calculations within the LSS approach \cite{LSS63}. The $SP_{n}$ values calculated using different approximations are also shown by different symbols for the references (see the text for details).}
\end{center}
\end{figure}

Figure~\ref{SPeLSScomp} shows the same $SP_{e}$ data \cite{Fastrup66,Hvelp68,Lennard86} (see Fig.~\ref{SPeSRIMcomp}) but in comparison to the calculations according to the LSS approach, in which reduced electronic stopping power $S_{e}^{\rm LSS}$ is determined as
\begin{equation}
  S_{e}^{\rm LSS} = K \varepsilon^{1/2},                           \label{SeLSS}
\end{equation}
where $K$ is given by
\begin{equation}
  K = \frac{0.0793 Z_{\rm HI}^{2/3} Z_{t}^{1/2} (A_{\rm HI}+A_{t})^{3/2}}{A_{\rm HI}^{3/2} A_{t}^{1/2} (Z_{\rm HI}^{2/3}+Z_{t}^{2/3})^{3/4}},                                          \label{kLSS}
\end{equation}
and reduced energy $\varepsilon$ is connected with HI energy $E$ (in keV) with the expression:
\begin{equation}
  \varepsilon = \frac{32.53A_{t}E}{(A_{\rm HI}+A_{t})Z_{\rm HI}Z_{t}(Z_{\rm HI}^{2/3}+Z_{t}^{2/3})^{1/2}},   \label{epsE}
\end{equation}
where $Z_{\rm HI}$, $Z_{t}$ are the HI and target atomic numbers, and $A_{\rm HI}$ and $A_{t}$ are the HI and target masses in amu.

Possible contributions of the nuclear stopping are also shown in the figure, which are calculated in different approximations for the reduced $S_{n}(\varepsilon)$ function. The least $SP_{n}$ values correspond to the $S_{n}(\varepsilon)$ approximation proposed by Ziegler \cite{Ziegler77} (designated as LSSZ in the figure):
\begin{eqnarray}
  S_{n }  & = &  1.7 \varepsilon^{1/2} \frac{\ln[\varepsilon + \exp(1)]}{1 + 6.8\varepsilon + 3.4\varepsilon^{3/2}}, \phantom{1.7} 0.01\leqslant\varepsilon \leqslant 10;                                 \nonumber             \\
  S_{n }  & = &  0.5 \ln(0.47\varepsilon)/\varepsilon,\phantom{\varepsilon^{1/2}0.5         \ln(0.47\varepsilon)/\varepsilon} \varepsilon > 10.          \label{SnLSSZ}
\end{eqnarray}
Other $SP_{n}$ curves correspond to $S_{n}(\varepsilon)$ obtained by the integration of the $f(\eta)$ function using Thomas-Fermi screening (see details, for example, in \cite{Sigmund04}) and by the approximation to the Kr-C free electron potential \cite{Wilson77} (designated as TF and Kr-C, respectively, in the figure). The Kr-C reduced $S_{n}(\varepsilon)$ function has a form:
\begin{equation}
  S_{n}^{\rm Kr-C} = 0.5 \ln(1 + \varepsilon)/(\varepsilon + 0.10718\varepsilon^{0.37544}).                        \label{SnWHB}
\end{equation}

The reduced stopping power values ($S_{e}^{\rm LSS}$ and $S_{n }$) are converted to the $SP_{e}^{\rm LSS}$ and $SP_{n }$ values in MeV/(mg/cm$^{2}$) with the relationship:
\begin{equation}
  SP_{e/n} = S_{e/n} \frac{5.0958 Z_{\rm HI} Z_{t} A_{\rm HI}}{A_{t}(A_{\rm HI}+A_{t})( Z_{\rm HI}^{2/3} + Z_{t}^{2/3})^{1/2}}.                                                                                                                                        \label{SntoSPn}
\end{equation}

\begin{figure}[!hb]  
\begin{center}
\includegraphics[width=0.485\textwidth]{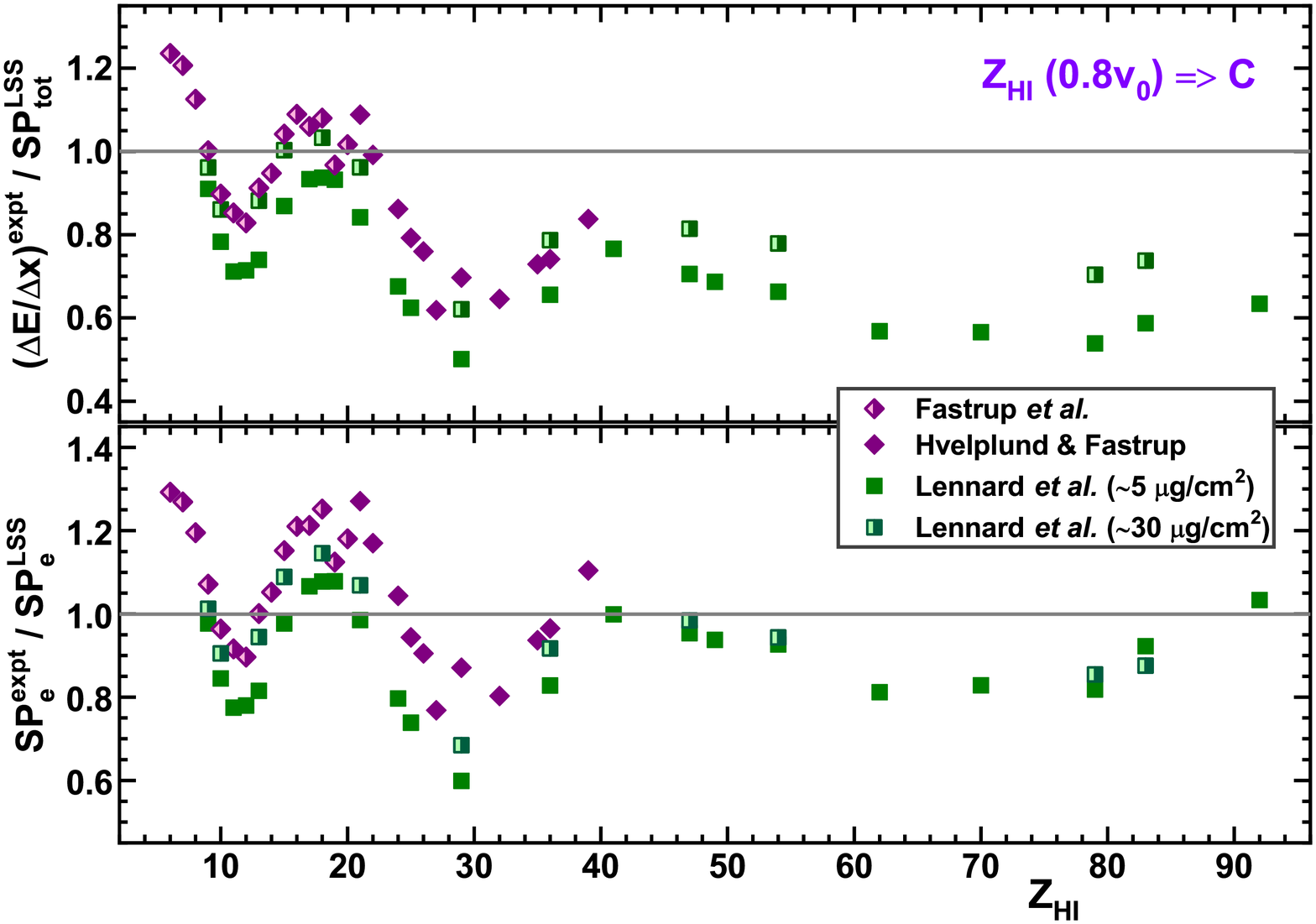}
\caption{\label{dEdXSPecomLSS} The same as in Fig.~\ref{dEdXSPecomp}, but for the $\Delta E/\Delta X$ data \cite{Fastrup66,Hvelp68,Lennard86} in comparison with the LSS calculations using Eqs.~(\ref{SeLSS})--(\ref{epsE}) for electronic stopping and (\ref{SnLSSZ}) for nuclear stopping (the upper panel). The comparison of the $SP_{e}$ values extracted in the same works and those calculated within the LSS approach is shown in the bottom panel. Both the comparisons to the experimental data were made using Eq.~(\ref{SntoSPn}).}
\end{center}
\end{figure}

Comparing Figs.~\ref{SPeSRIMcomp} and \ref{SPeLSScomp}, one can see that $SP_{n}$ values calculated with  Eqs.~(\ref{SnLSSZ}) and (\ref{SntoSPn}) give us the least contribution to the total SP values. In Fig.~\ref{dEdXSPecomLSS}, the $SP_{e}$ and $\Delta E/\Delta X$ values obtained in the experiments  \cite{Fastrup66,Hvelp68,Lennard86} are compared with the respective $SP_{e}^{\rm LSS}$ and $SP_{tot}^{\rm LSS}$ calculated values. This figure could be compared with Fig.~\ref{dEdXSPecomp} showing similar SP ratios according to SRIM calculations. As expected and seen in Fig.~\ref{dEdXSPecomLSS}, the calculations do not reproduce oscillations in $SP_{e}$ for $Z_{\rm HI} \lesssim 40$, although LSS calculations are closer to the $SP_{e}$ data for $Z_{\rm HI} \gtrsim 40$ than those obtained with SRIM. At the same time, the contributions of the nuclear stopping obtained with Eqs.~(\ref{SnLSSZ}) and (\ref{SntoSPn}) are still overestimated, as it follows from the upper panel of the figure.

In the next section, available data on total and electronic stopping powers are considered as a function of the HI velocity and compared with the respective values obtained within SRIM calculations. Within this consideration, attempts are made to estimate the nuclear stopping power from such a comparison.

\section{\label{totSP}Stopping power velocity dependence}

As was mentioned above, SP data obtained in many experiments relate to the total SP values. At relatively high velocities, when nuclear stopping is negligible, these data may refer to the electronic SP component.  In the low energy SP measurements \cite{Fastrup66,Hvelp68,Lennard86}, the experiments were conducted within the narrow range of forward angles for HIs escaping the targets. In contrast to the $SP_{n}$ component, the $SP_{e}$ component is independent of the escaping angle. Thus, the $SP_{n}$ values considered below refer to the ``forward-direction'' ones. In this sense, the results of the analysis \cite{Garnir80} mentioned above showed two distinguished empirical approximations for reduced nuclear stopping, which corresponded to the ``all angles'' and ``forward-direction'' values.

\begin{figure}[!h]  
\begin{center}
\includegraphics[width=0.485\textwidth]{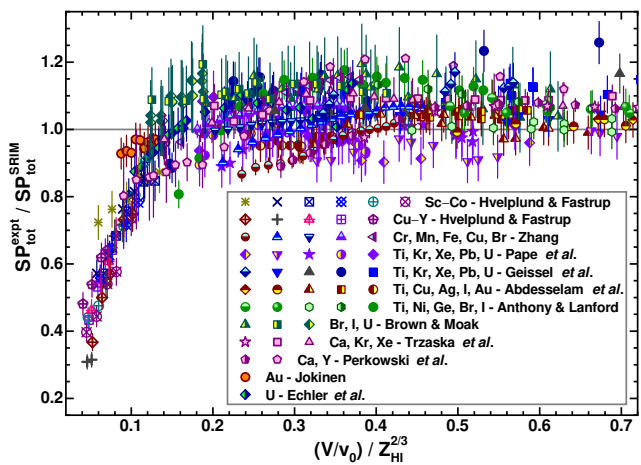}
\caption{\label{SPtallSRIM}The ratios of total SP values obtained in experiments $SP_{tot}^{\rm expt}$ \cite{SRIM,IAEASP} and those calculated with SRIM $SP_{tot}^{\rm SRIM}$ \cite{SRIM} are shown as a function of HI reduced velocity $(V/v_{0}) / Z^{2/3}_{\rm HI}$ (different symbols corresponding to the indicated HIs). The references to the first author of the works are also indicated.}
\end{center}
\end{figure}

The ratios of total SP values obtained in the experiments $SP_{tot}^{\rm expt}$ \cite{SRIM,IAEASP} and those calculated with SRIM $SP_{tot}^{\rm SRIM}$ \cite{SRIM} give us a general trend in the total SP dependency on the HI velocity. Figure~\ref{SPtallSRIM} shows these ratios as a function of reduced velocity $V_{r} = (V/v_{0}) / Z^{2/3}_{\rm HI}$ for $Z_{\rm HI} \geqslant 20$ ($V$ and $v_{0}$ are HI and Bohr velocities, respectively). The $V_{r}$ parameterization was earlier used for the derivation of HI effective charges from SP data in many works (see, for example,  \cite{BrownMoak72,Pape78,AnthLanf82,SchulzBrandt82,Abdess92,Saga2015}).

As one can see in the figure, at $V_{r} \gtrsim 0.2$ SRIM reproduces $SP_{tot}^{\rm expt}$ data in general with some deviations within the range of +20 to -15\%. At $V_{r} \lesssim 0.15$, SRIM overestimates the $SP_{tot}^{\rm expt}$ values for all HIs under analysis. These inconsistencies could not be explained by the $SP_{e}^{\rm SRIM}$ overestimates (see Fig.~\ref{SPeSRIMcomp} and the bottom panel in Fig.~\ref{dEdXSPecomp}), but are the result of the respective overestimates in the $SP_{n}^{\rm SRIM}$ values. Specific deviations of the experimental data from the calculations could be caused by specific $Z_{\rm HI}$ stopping that could differ from those predicted by SRIM. This circumstance requires SP data considerations for every specified ion.

\subsection{\label{SPtotSRIM}Total stopping power data analysis}

\begin{figure}[!h]   
\begin{center}
\includegraphics[width=0.485\textwidth]{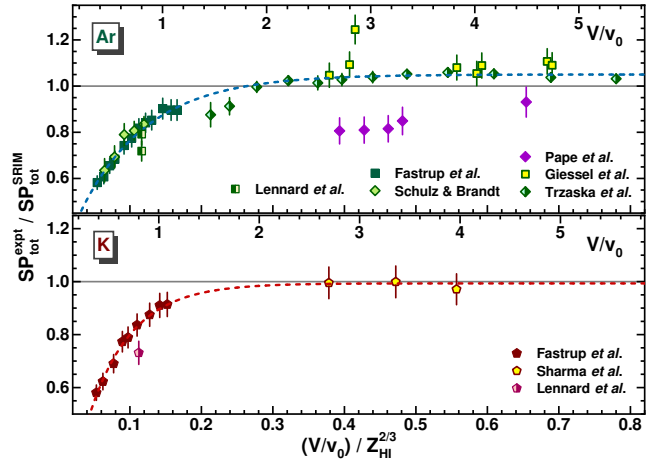}
\caption{\label{ArKSPtSRIM}The same as in Fig.~\ref{SPtallSRIM}, but for Ar and K ions only (upper and bottom panels, respectively). The results of data fits with Eq.~(\ref{expfit}) are shown by dashed lines (the data \cite{Lennard86,Pape78} were excluded as mentioned in the text). For orientation, relative velocity $V/v_{0}$ is shown in the upper axes of the panels.}
\end{center}
\end{figure}

\begin{figure}[!h]   
\begin{center}
\includegraphics[width=0.485\textwidth]{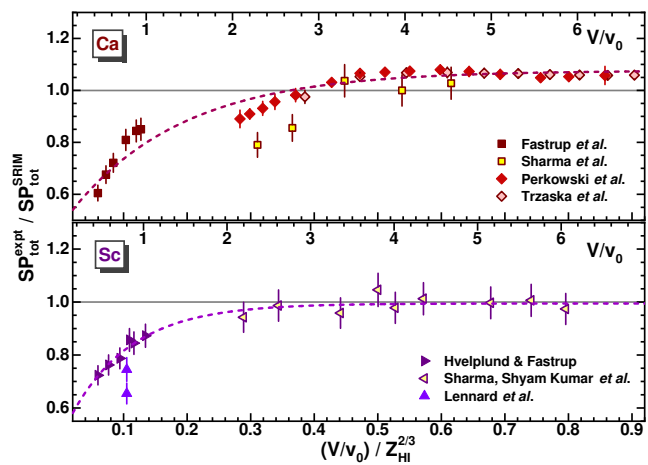}
\caption{\label{CaScSPtSRIM} The same as in Fig.~\ref{ArKSPtSRIM}, but for Ca and Sc ions (upper and bottom panels, respectively). See the text for details.}
\end{center}
\end{figure}

\begin{figure}[!h]   
\begin{center}
\includegraphics[width=0.485\textwidth]{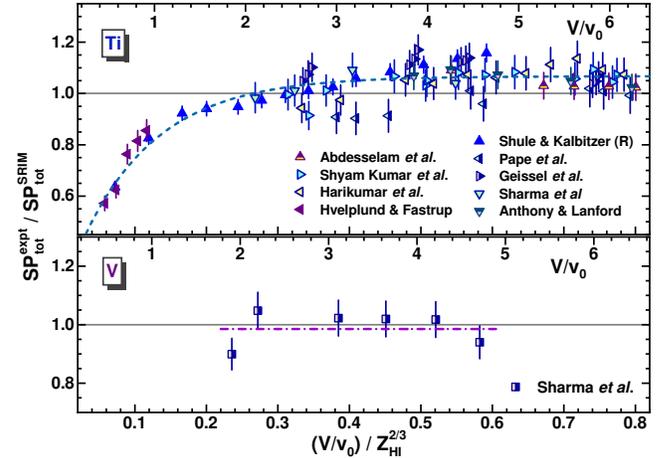}
\caption{\label{TiVSPtSRIM} The same as in Figs.~\ref{ArKSPtSRIM} and \ref{CaScSPtSRIM}, but for Ti and V ions (upper and bottom panels, respectively). See the text for details.}
\end{center}
\end{figure}

\begin{figure}[!h]   
\begin{center}
\includegraphics[width=0.485\textwidth]{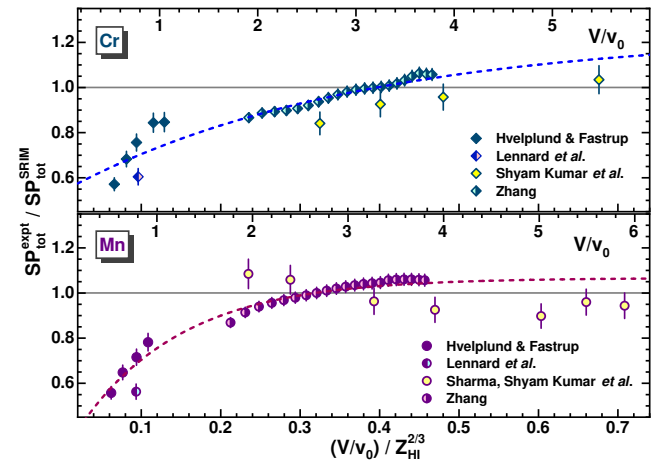}
\caption{\label{CrMnSPtSRIM} The same as in Figs.~\ref{ArKSPtSRIM}--\ref{TiVSPtSRIM}, but for Cr and Mn ions (upper and bottom panels, respectively). See the text for details.}
\end{center}
\end{figure}

\begin{figure}[!h]   
\begin{center}
\includegraphics[width=0.485\textwidth]{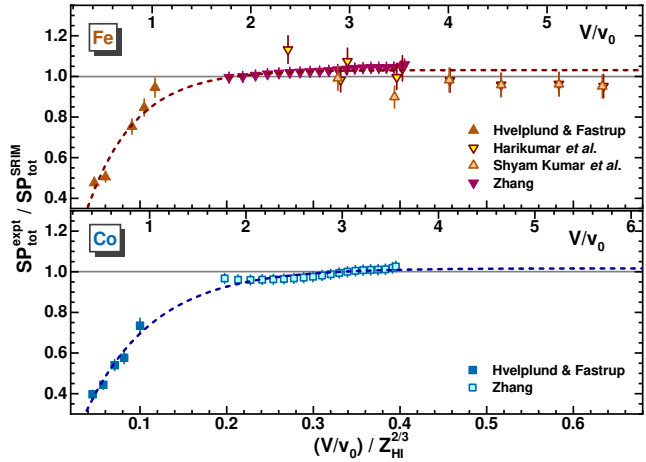}
\caption{\label{FeCoSPtSRIM} The same as in Figs.~\ref{ArKSPtSRIM}--\ref{CrMnSPtSRIM}, but for Fe and Co ions (upper and bottom panels, respectively). See the text for details.}
\end{center}
\end{figure}

\begin{figure}[!h]   
\begin{center}
\includegraphics[width=0.485\textwidth]{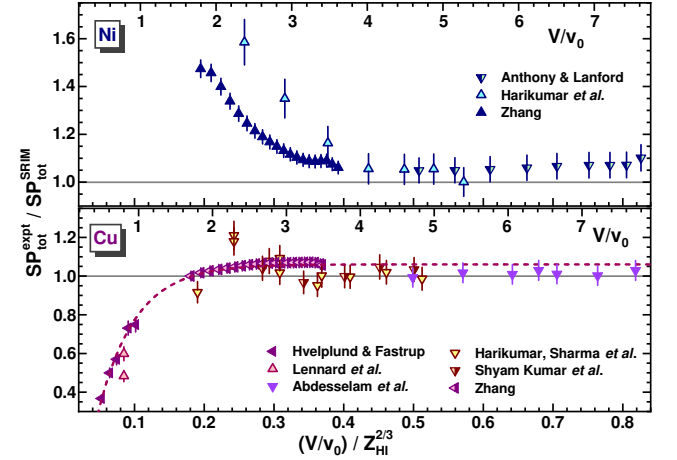}
\caption{\label{NiCuSPtSRIM} The same as in Figs.~\ref{ArKSPtSRIM}--\ref{CrMnSPtSRIM}, but for Ni and Cu ions (upper and bottom panels, respectively). See the text for details.}
\end{center}
\end{figure}

\begin{figure}[!h]   
\begin{center}
\includegraphics[width=0.485\textwidth]{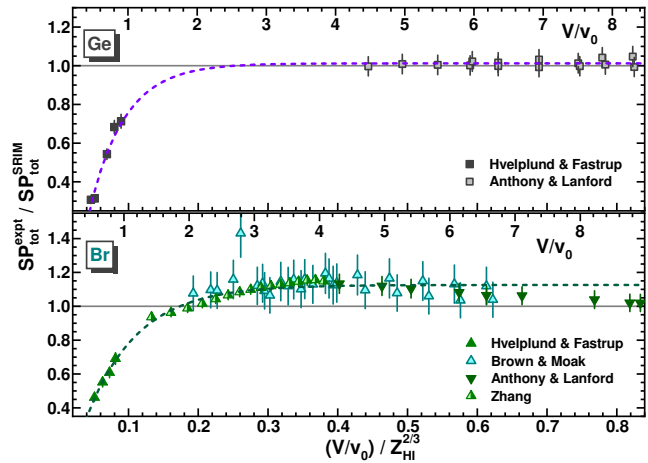}
\caption{\label{GeBrSPtSRIM} The same as in Figs.~\ref{ArKSPtSRIM}--\ref{NiCuSPtSRIM}, but for Ge and Br ions (upper and bottom panels, respectively). See the text for details.}
\end{center}
\end{figure}

\begin{figure}[!h]   
\begin{center}
\includegraphics[width=0.485\textwidth]{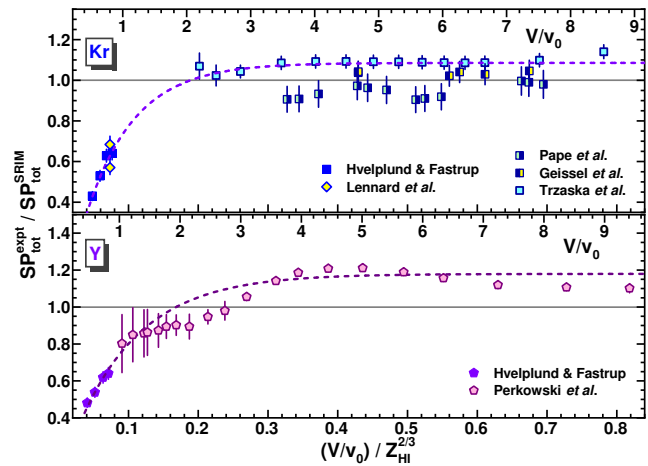}
\caption{\label{KrYSPtSRIM} The same as in Figs.~\ref{ArKSPtSRIM}--\ref{GeBrSPtSRIM}, but for Kr and Y ions (upper and bottom panels, respectively). See the text for details.}
\end{center}
\end{figure}

\begin{figure}[!h]   
\begin{center}
\includegraphics[width=0.485\textwidth]{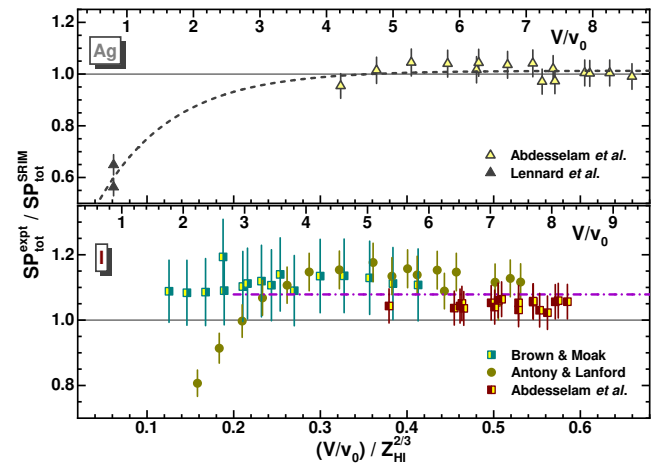}
\caption{\label{AgISPtSRIM} The same as in Figs.~\ref{ArKSPtSRIM}--\ref{KrYSPtSRIM}, but for Ag and I ions (upper and bottom panels, respectively). See the text for details.}
\end{center}
\end{figure}

\begin{figure}[!h]   
\begin{center}
\includegraphics[width=0.485\textwidth]{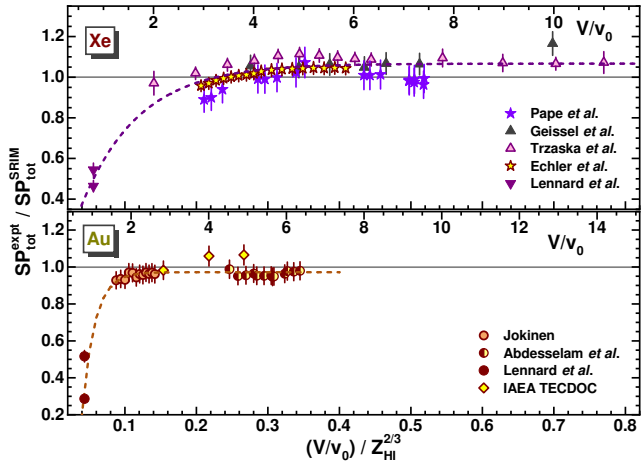}
\caption{\label{XeAuSPtSRIM} The same as in Figs.~\ref{ArKSPtSRIM}--\ref{AgISPtSRIM}, but for Xe and Au ions (upper and bottom panels, respectively). See the text for details.}
\end{center}
\end{figure}

\begin{figure}[!h]  
\begin{center}
\includegraphics[width=0.485\textwidth]{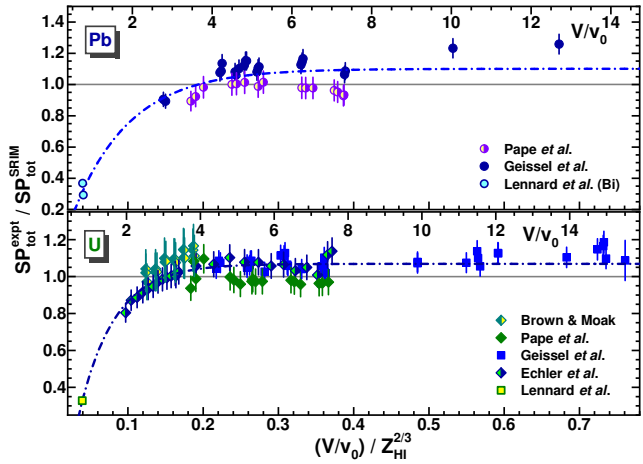}
\caption{\label{PbUSPtSRIM}The same as in Figs.~\ref{ArKSPtSRIM}--\ref{XeAuSPtSRIM}, but for Pb and U ions only (upper and bottom panels, respectively). See the text for details.}
\end{center}
\end{figure}

In Figs.~\ref{ArKSPtSRIM}--\ref{PbUSPtSRIM}, the ratios of total SP values obtained in experiments and those obtained in SRIM calculations are shown for HIs from Ar to U. Some details on the $SP_{tot}^{\rm expt}$ data used in the subsequent analysis should be mentioned. Available original data (with original errors) were preferably used in the analysis. In the absence of tabulated data in the original works, SP data from the database \cite{IAEASP} were used.

The $SP_{tot}^{\rm expt}/SP_{tot}^{\rm SRIM}$ values as a function of $V_{r}$ for Ar to U ions were fitted using the correction function in the framework of the weighted LSM procedure:
\begin{equation}  \label{expfit}
 f_{cor}(V_{r}) \equiv SP_{tot}^{\rm expt}/SP_{tot}^{\rm SRIM} = a - b\exp(-k V_{r}),
\end{equation}
where $a$, $b$, and $k$ are fitting parameters.

The ratios of the $SP_{tot}^{\rm expt}$ and $SP_{tot}^{\rm SRIM}$ values for Ar and K ions are shown in Fig.~\ref{ArKSPtSRIM}. The Ar and K $SP_{tot}^{\rm expt}$ data were taken from the respective tables \cite{Fastrup66,Lennard86,Pape78,Trzaska18,Sharma99}, whereas the Ar data obtained by Giessel \ea\ were taken from the database \cite{IAEASP}. The low velocity Ar SP data \cite{SchulzBrandt82} were obtained using tabulated input and output energies ($E_{in}$ and $E_{out}$, respectively), target thickness $\Delta X$, and the relationship:
\begin{equation}
  SP_{tot}^{\rm expt} = (E_{in} - E_{out}) / \Delta X.                                                                                 \label{dEdX}
\end{equation}
As one can see in the figure, the Ar data obtained recently \cite{Trzaska18} and earlier by Giessel \ea\ \cite{IAEASP} are in agreement with each other, whereas the data \cite{Pape78} at $V_{r} \gtrsim 0.4$ are inconsistent with them.  The  Ar and K data \cite{Lennard86} for a thin target at $0.8 v_{0}$ is also in disagreement with those of \cite{Fastrup66,SchulzBrandt82}. The data \cite{Lennard86,Pape78} were excluded from the data fit. Fitting parameter values thus obtained for the Ar and K data are listed in Table~\ref{abkparam}. As one could expect, adding the data \cite{Lennard86,Pape78} led to an increase in fitted parameter errors and in the reduced $\chi^{2}_{r}$ value determining fit quality.

Figure~\ref{CaScSPtSRIM} shows the Ca and Sc SP data \cite{Fastrup66,Trzaska18,Sharma99,Perkowski06,ShyKum96} in comparison to SRIM calculations. As one can see, the fitting curve does not provide the match of the Ca data \cite{Fastrup66,Trzaska18,Perkowski06} using the function determined by Eq.~(\ref{expfit}) even with omitting the data \cite{Sharma99} at $0.3 < V_{r} < 0.4$. It is implied that the data of \cite{Fastrup66} at relatively high velocities and those of \cite{Perkowski06} at relatively low velocities are reliable. In this regard, it is worth noting that the reduced $\chi^{2}_{r}$ value obtained for the Ca data fit (1.66) is greater than the one obtained for Sc (0.18) indicating a good match, if the data \cite{Lennard86} were ignored. The last correspond to the noticeably lower SP values than those obtained in \cite{Fastrup66}.

Figure~\ref{TiVSPtSRIM} shows the Ti and V SP data in comparison to SRIM calculations. The Ti data at relatively high velocities ($V_{r} \gtrsim 0.3$) \cite{Pape78,AnthLanf82,Abdess92,Sharma99,ShyKum96,Harikumar96} and those obtained by Giessel \ea\ \cite{IAEASP} are in satisfactory agreement with each other (the data \cite{AnthLanf82} were taken from the database \cite{IAEASP}).  The data from range measurements \cite{ShuKal91} (designated by R in the figure) agree with the low velocity data \cite{Hvelp68} and those obtained at relatively high velocities. These data were originally assigned to the electronic SP \cite{ShuKal91} and were taken from the database \cite{IAEASP}. They were attributed to the total SP (it seems impossible to separate the inelastic and elastic components in range measurements considering the dominance of elastic collisions at the end of the range).  The V data \cite{Sharma99} were fitted with a constant value for the $SP_{tot}^{\rm expt}/SP_{tot}^{\rm SRIM}$ ratios. The results of data fits for the Ti and V ions are listed in Table~\ref{abkparam}.

Figure~\ref{CrMnSPtSRIM}  shows a comparison of the Cr and Mn SP data \cite{Zhang2002,Hvelp68,Lennard86,ShyKum96,Sharma99} to SRIM calculations  (the data \cite{Zhang2002} were taken from the database \cite{IAEASP}). The Cr data at $0.3 \lesssim V_{r} \lesssim 0.5$ \cite{ShyKum96} and the Mn data at $0.2 \lesssim V_{r} \lesssim 0.5$ \cite{Sharma99} are in some disagreement with those obtained with small errors \cite{Zhang2002} later. The fitting curve does not provide the match of the Cr data \cite{Zhang2002} with those at low velocities \cite{Hvelp68} (similarly to the Ca data). The reason for this is that in the fitting procedure the precision data \cite{Zhang2002} contributed more weights than the less precise \cite{Hvelp68,ShyKum96,Sharma99}. The data at $0.8v_{0}$ \cite{Lennard86} disagree with the data \cite{Hvelp68} at the same velocity, as shown in the figure, and were ignored in the fitting procedure. The results of fitting are listed in Table~\ref{abkparam}.

\begin{table}[!h]  
\vspace*{-1.5mm}
\caption{\label{abkparam} Fitting parameter values $a$, $b$, and $k$ as obtained for Ar to U ratios $SP_{tot}^{\rm expt}/SP_{tot}^{\rm SRIM}$ fitted with Eq.~(\ref{expfit}). The ion symbols and ranges of the applicability of Eq.~(\ref{expfit}) for specified ions are listed in the first and last columns, respectively. The results of data fitting are also shown in Figs.~\ref{ArKSPtSRIM}--\ref{PbUSPtSRIM}.
}
\begin{ruledtabular}
\begin{tabular}{lccrr}
Ion                                   &                    $a$               &                $b$        & \multicolumn{1}{c}{$k$} &$V_{r}$ range \\
\hline
  Ar\footnotemark[1] & 1.0502$\pm$0.0063 & 0.829$\pm$0.061 & 10.62$\pm$0.94 &  0.05--0.8   \\
   K\footnotemark[2]  & 0.9934$\pm$0.0105 & 1.027$\pm$0.055 & 16.71$\pm$1.05 &  0.05--0.6   \\
 Ca\footnotemark[3]  & 1.0764$\pm$0.0087 & 0.601$\pm$0.045 &   5.69$\pm$0.67 &   0.06--0.9   \\
  Sc\footnotemark[2]  & 0.9946$\pm$0.0095 & 0.511$\pm$0.062 & 10.49$\pm$1.60 &   0.06--0.8  \\
  Ti                                     & 1.0672$\pm$0.0085 & 0.827$\pm$0.061 &   9.46$\pm$0.96 &  0.06--0.8    \\
  V                                      & 0.9852$\pm$0.0274 &                                     &                                   &    0.2--0.6    \\
  Cr\footnotemark[2]  & 1.2383$\pm$0.1157 & 0.700$\pm$0.089 &   2.72$\pm$0.92 &    0.06--0.7  \\
 Mn\footnotemark[2] & 1.0664$\pm$0.0193 & 0.786$\pm$0.064 &   7.75$\pm$1.14 &    0.06--0.7  \\
  Fe                                     & 1.0311$\pm$0.0065 & 1.380$\pm$0.149 & 18.02$\pm$1.96 &   0.04--0.7  \\
  Co                                    & 1.0167$\pm$0.0074 & 1.140$\pm$0.050 & 12.69$\pm$0.88 &   0.04--0.5  \\
  Cu\footnotemark[2]  & 1.0605$\pm$0.0058 & 1.874$\pm$0.144 & 18.75$\pm$1.32 &   0.05--0.8  \\
  Ge                                    & 1.0122$\pm$0.0117 & 1.827$\pm$0.142 & 19.81$\pm$1.56 &   0.04--0.9  \\
  Br                                     & 1.1261$\pm$0.0081 & 1.352$\pm$0.070 & 13.67$\pm$0.86 &   0.04--0.9  \\
  Kr \footnotemark[4]  & 1.0856$\pm$0.0075 & 1.326$\pm$0.115 & 14.51$\pm$1.49 &   0.04--0.8  \\
   Y                                      & 1.1794$\pm$0.0141 & 1.127$\pm$0.133 & 10.92$\pm$2.08  &   0.04--0.8  \\
  Ag                                    & 1.0129$\pm$0.0130 & 0.847$\pm$0.332 & 11.70$\pm$6.67  &   0.06--0.7  \\
    I                                     & 1.0786$\pm$0.0069 &                                      &                                    &    0.2--0.6    \\
  Xe                                    & 1.0668$\pm$0.0090 & 1.140$\pm$0.083 & 12.46$\pm$1.19  &   0.05--0.8   \\
  Au                                    & 0.9718$\pm$0.0174 & 7.44$\pm$7.48      & 56.8$\pm$23.4    &   0.04--0.4   \\
  Pb                                    & 1.1008$\pm$0.0263 & 1.335$\pm$0.105  & 13.38$\pm$2.40  &   0.04--0.8  \\
   U                                     & 1.0687$\pm$0.0067 & 1.711$\pm$0.102  & 21.28$\pm$1.42  &   0.04--0.8  \\
\end{tabular}
\end{ruledtabular}
\footnotetext[1]{Without the data \cite{Lennard86,Pape78} (see the text).}
\footnotetext[2]{Without the data \cite{Lennard86} (see the text).}
\footnotetext[3]{Without the data \cite{Sharma99} at $0.3 < V_{r} < 0.4$ (see the text)}.
\footnotetext[4]{Without the data \cite{Pape78} (see the text).}
\end{table}

Figure~\ref{FeCoSPtSRIM} shows the Fe and Co SP data \cite{Zhang2002,Hvelp68,ShyKum96,Harikumar96,Harikumar97} in comparison to SRIM calculations. In contrast to the Ca and Cr data analysis (see Figs.~\ref{CaScSPtSRIM} and \ref{CrMnSPtSRIM}), fitting curves provide an acceptable match of the Fe and Co data at $V_{r} \gtrsim 0.2$ \cite{Zhang2002} (taken from the database \cite{IAEASP}) with those at low velocities \cite{Hvelp68} ($\chi^{2}_{r} = 0.985$ and 0.546 for Fe and Co data fitting, respectively). The results of fitting are listed in Table~\ref{abkparam}.

Figure~\ref{NiCuSPtSRIM} shows the Ni and Cu SP data \cite{Zhang2002,Hvelp68,AnthLanf82,Abdess92,Sharma99,ShyKum96,Harikumar96,Harikumar97} in comparison to SRIM calculations. The Ni data \cite{AnthLanf82,Harikumar97} at $V_{r} > 0.4$ are in good agreement with each other and exceed SRIM calculations within $\sim$5\%. At $V_{r} < 0.4$, the data \cite{Zhang2002,Harikumar97} significantly exceed SRIM calculations, as seen in the figure (the data \cite{Zhang2002} were taken from the database \cite{IAEASP}). Thus, the Ni data were not processed using Eq.~(\ref{expfit}). As for Cu data, the fitting curve provides the match of the high velocity data \cite{Zhang2002,Hvelp68,AnthLanf82,Abdess92,Sharma99,ShyKum96,Harikumar97} with the low velocity ones \cite{Hvelp68}. Ignoring the data \cite{Lennard86} the least $\chi^{2}_{r} = 0.774$ value was obtained. The results of fitting are listed in Table~\ref{abkparam}.

Figure~\ref{GeBrSPtSRIM}  compares Ge and Br SP data \cite{Zhang2002,Hvelp68,BrownMoak72,AnthLanf82} to SRIM calculations  (the data \cite{Zhang2002} were taken from the database \cite{IAEASP}). A lack of data for Ge ions at middle velocities ($0.1 \lesssim V_{r} \lesssim 0.4$) makes the fitting parameters ($b$ and $k$ values) somewhat questionable, despite a good data fit ($\chi^{2}_{r} = 0.809$). The Br data at $V_{r} \lesssim 0.6$ \cite{Zhang2002,BrownMoak72,AnthLanf82} are reasonably consistent. The data \cite{BrownMoak72} presented as the $SP_{e}$ values were corrected for $SP_{n}$. The last was calculated with an approximate expression given in reduced values \cite{LNS68}:  $S_{n}(\varepsilon) = 0.5\ln(1.294\varepsilon)/\varepsilon$. This correction corresponded to $\simeq$1.5\% of the $SP_{tot}$ value at the lowest velocities of the data \cite{BrownMoak72} given with 10\% accuracy.  The results of fitting are listed in Table~\ref{abkparam}.

Figure~\ref{KrYSPtSRIM} shows the Kr and Y SP data \cite{Hvelp68,Lennard86,Pape78,Trzaska18,Perkowski09} and those obtained by Geissel \ea\ \cite{IAEASP} in comparison to SRIM calculations. As in the case of Ar data, the Kr data \cite{Pape78} at $V_{r}\gtrsim 0.3$ lie noticeably below the data of Geissel \ea\ and \cite{Trzaska18}, which are in satisfactory agreement with each other. The last two together with the low velocity data \cite{Hvelp68,Lennard86} are well fitted with Eq.~(\ref{expfit}), as shown in the figure ($\chi^{2}_{r} = 0.669$). Despite well matching the Y data \cite{Perkowski09} to the low velocity data \cite{Hvelp68}, the data fit yielded a large value of  $\chi^{2}_{r} = 7.38$. Implying the data \cite{Perkowski09} reliability, a bad data fit could be explained by a simplified fitting model using Eq.~(\ref{expfit}), which is unable to describe the data with small errors at $V_{r} \gtrsim 0.2$.

Figure~\ref{AgISPtSRIM} shows the Ag and I SP data \cite{Lennard86,BrownMoak72,AnthLanf82,Abdess92} in comparison to SRIM calculations. The Ag data obtained in \cite{Lennard86,Abdess92} are the only one available. The data \cite{Abdess92} were treated as the $SP_{tot}^{\rm expt}$ values because nuclear stopping contributes only 0.8\% of the total stopping at the lowest velocity, according to SRIM calculations. This value is much lower than the 5\% total uncertainty assigned to the data. The I data \cite{BrownMoak72,AnthLanf82,Abdess92} at $V_{r} > 0.2$ are quite agreeable with each other, whereas the data \cite{BrownMoak72,AnthLanf82} are varied at $V_{r} < 0.2$. The $SP_{e}^{\rm expt}$ data \cite{BrownMoak72} were corrected for $SP_{n}$ in the same way as the Br data \cite{BrownMoak72} were. Eq.~(\ref{expfit}) was used to fit the Ag data, whereas the I data at $V_{r} > 0.2$ could be fitted with a constant for the $SP_{tot}^{\rm expt}/SP_{tot}^{\rm SRIM}$ ratios, as was done for the V data (see Fig.~\ref{TiVSPtSRIM}).

Figure~\ref{XeAuSPtSRIM} shows the Xe and Au SP data \cite{Lennard86,Pape78,Abdess92,Trzaska18,Echler17,Jokinen97}, and those obtained by Geissel \ea\ for Xe, and indicated as IAEA TECDOC for Au (both were taken from the database \cite{IAEASP}), which are compared to SRIM calculations. The Xe data at $0.2 < V_{r} < 0.6$ are close to each other. The data \cite{Echler17} originally presented as $SP_{e}^{\rm expt}$ were limited to $V_{r} \gtrsim 0.2$. At these velocities, $SP_{n}^{\rm SRIM}$ values contribute less than 3\% of $SP_{tot}^{\rm SRIM}$, which is less than the respective data errors assigned in \cite{Echler17}. The Au data \cite{Abdess92} were assigned to the $SP_{tot}^{\rm expt}$ values because nuclear stopping contributes only 1.5\% of total stopping, according to SRIM calculations at the lowest energy. This contribution is much less than the respective errors assigned in the work. For the Au data \cite{Jokinen97}, $SP_{tot}^{\rm expt} = \Delta E/\Delta X$ estimates were based on the tabulated $\Delta E$ data. These data corresponded to the Au input energies $E_{in} = 15-37$ MeV. The average energies for the $SP_{tot}^{\rm expt}$ values thus obtained were estimated as $E_{av} = E_{in} - SP_{tot}^{\rm expt} \Delta X/2$. The Xe and Au data at $0.8 v_{0}$ \cite{Lennard86}, corresponding to the different target thickness, were added, and both the data sets were fitted with Eq.~(\ref{expfit}). The steep fall in the Au SP data in going from $0.08\lesssim V_{r} \lesssim 0.14$ \cite{Jokinen97} to those at $0.8 v_{0}$ \cite{Lennard86} led to the large $b$ and $k$ fitted values obtained with large errors. These values noticeably exceeded those obtained for other HIs (see Table~\ref{abkparam}).

Figure~\ref{PbUSPtSRIM} shows the Pb and U SP data \cite{Lennard86,BrownMoak72,Pape78,Echler12}, and those obtained by Geissel \ea\ \cite{IAEASP} in comparison to SRIM calculations. Though the Pb data of Geissel \ea\ and \cite{Pape78} are in some disagreement with each other, they were fitted with Eq.~(\ref{expfit}) altogether. A similar difference is seen for the U data of the same authors. These data, together with others \cite{BrownMoak72,Echler12}, are in satisfactory agreement with each other. The U data \cite{BrownMoak72} presented as the $SP_{e}$ values were corrected for $SP_{n}$ in the same way as was done for the Br and I data \cite{BrownMoak72}. The resulting $SP_{tot}^{\rm expt}$ data are slightly higher than those obtained later at the same velocities \cite{Echler12}, but they are consistent within error bars with the rest of the data, as seen in the figure. The Pb and U data were supplemented with the low velocity Bi and U data \cite{Lennard86}. In doing so, a possible distinction in the SP values for Pb and Bi was  neglected. The results of fitting are listed in Table~\ref{abkparam}.

\begin{figure}[!h]  
\begin{center}
\includegraphics[width=0.4\textwidth]{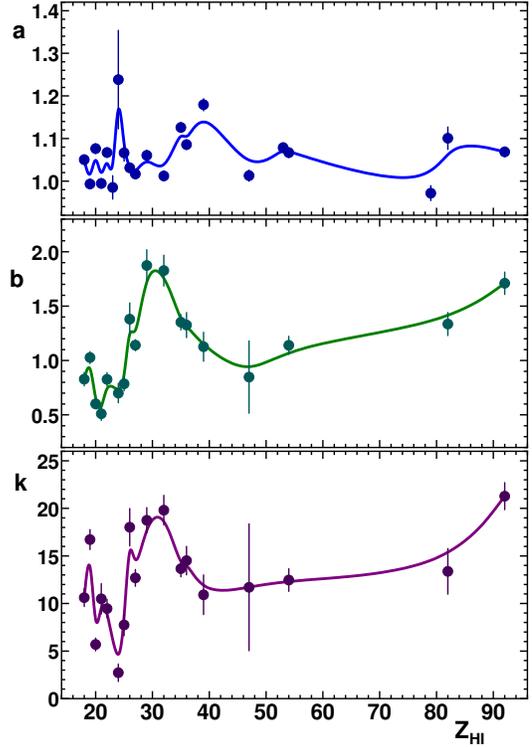}
\vspace*{-1.5mm}
\caption{\label{abkpar}Fitting parameters $a$, $b$, and $k$ listed in Table~\ref{abkparam} are shown by filled circles as functions of the HI atomic number (from upper to bottom panels, respectively). Solid lines are B-spline approximations of the data.}
\end{center}
\end{figure}

In Fig.~\ref{abkpar}, the fitting parameters listed in Table~\ref{abkparam} are shown as a function of the HI atomic number. As one can see in the figure, amplitude $a$, corresponding to the $SP_{tot}^{\rm expt}/SP_{tot}^{\rm SRIM}$ ratio at high velocities, oscillates in a sporadic way within a magnitude of about 0.9--1.2. The exponent parameters ($b$ and $k$), which determine decreasing $SP_{tot}^{\rm expt} / SP_{tot}^{\rm SRIM}$ ratios at low velocities, correlate with each other to a certain extent in the region of $24 \lesssim Z_{\rm HI} \lesssim 40$ and, probably, for higher $Z_{\rm HI}$ (the Au parameters have been omitted due to their large errors).

Now, using Eqs.~(\ref{SPtot}) and (\ref{expfit}) with the parameter values listed in Table~\ref{abkparam}, one can estimate the $SP_{n}^{\rm expt}$ values as
\begin{equation}
  SP_{n}^{\rm expt} = f_{cor}(V_{r}) SP_{tot}^{\rm SRIM} - SP_{e}^{\rm expt},                                          \label{SPnexpt}
\end{equation}
where $SP_{e}^{\rm expt}$ are the electronic SPs obtained in experiments. The $SP_{e}^{\rm expt}$ estimates are considered in the next section, applying the same approximation of Eq.~(\ref{expfit}) to the available $SP_{e}^{\rm expt}$ data and $SP_{e}^{\rm SRIM}$ calculations as was done above for the $SP_{tot}^{\rm expt}$ data analysis.

\subsection{\label{SPeSRIM}Electronic stopping power data analysis}

\begin{figure}[!h]   
\begin{center}
\includegraphics[width=0.485\textwidth]{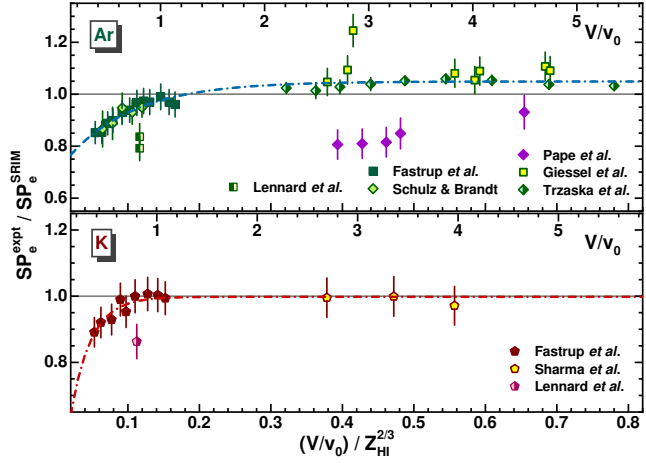}
\caption{\label{ArKSPeSRIM}$SP_{e}^{\rm expt}/SP_{e}^{\rm SRIM}$ ratios for Ar and K data \cite{Fastrup66,Lennard86,Pape78,Trzaska18,SchulzBrandt82,Sharma99} as well as Geissel \ea\ \cite{IAEASP} Ar data (upper and bottom panels, respectively). The results of the data fitting with Eq.~(\ref{expSPefit}) are shown by dash-dotted lines (the data \cite{Lennard86} and  \cite{Pape78} were excluded). The upper axes of the panels correspond to relative velocity $V/v_{0}$ shown for orientation.}
\end{center}
\end{figure}

\begin{figure}[!h]   
\begin{center}
\includegraphics[width=0.485\textwidth]{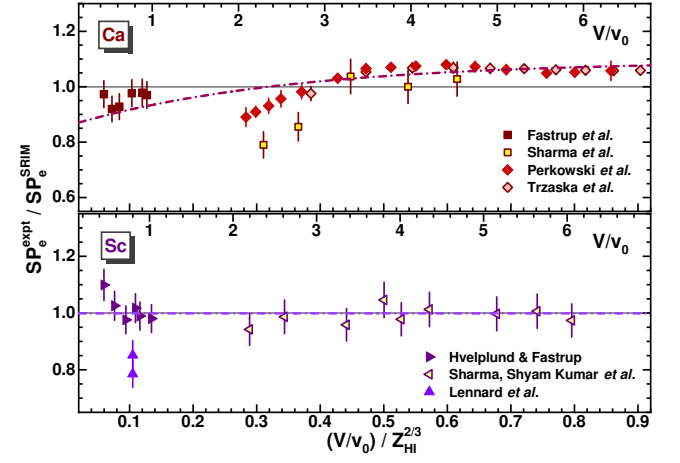}
\caption{\label{CaScSPeSRIM} The same as in Fig.~\ref{ArKSPeSRIM}, but for Ca and Sc ions (upper and bottom panels, respectively). See the text for details.}
\end{center}
\end{figure}

\begin{figure}[!h]   
\begin{center}
\includegraphics[width=0.485\textwidth]{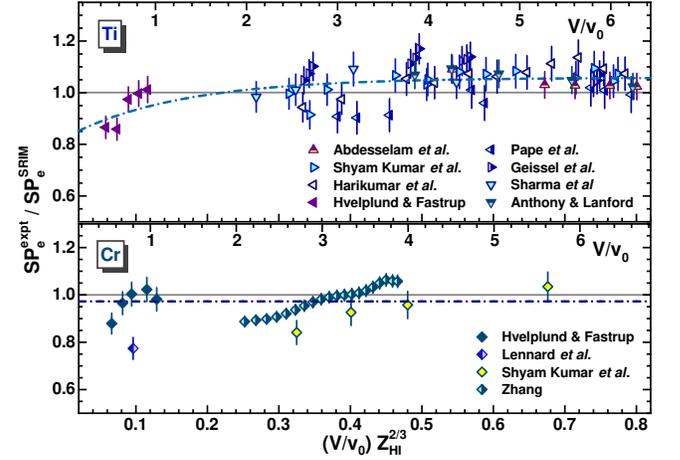}
\caption{\label{TiCrSPeSRIM} The same as in Figs.~\ref{ArKSPeSRIM} and \ref{CaScSPeSRIM}, but for Ti and Cr ions (upper and bottom panels, respectively). See the text for details.}
\end{center}
\end{figure}

\begin{figure}[!h]   
\begin{center}
\includegraphics[width=0.485\textwidth]{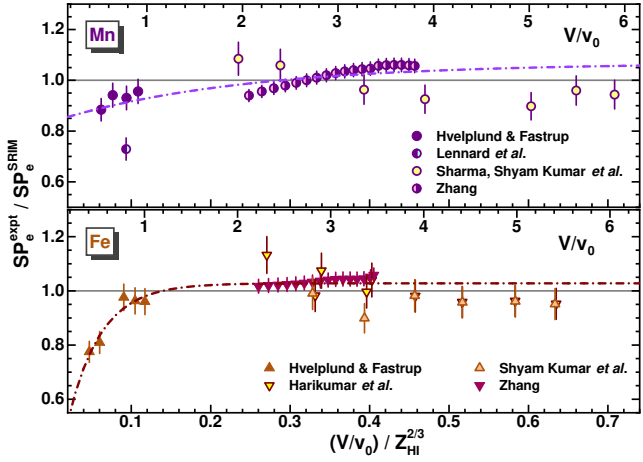}
\caption{\label{MnFeSPeSRIM} The same as in Figs.~\ref{ArKSPeSRIM}--\ref{TiCrSPeSRIM}, but for Mn and Fe ions (upper and bottom panels, respectively). See the text for details.}
\end{center}
\end{figure}

\begin{figure}[!h]   
\begin{center}
\includegraphics[width=0.485\textwidth]{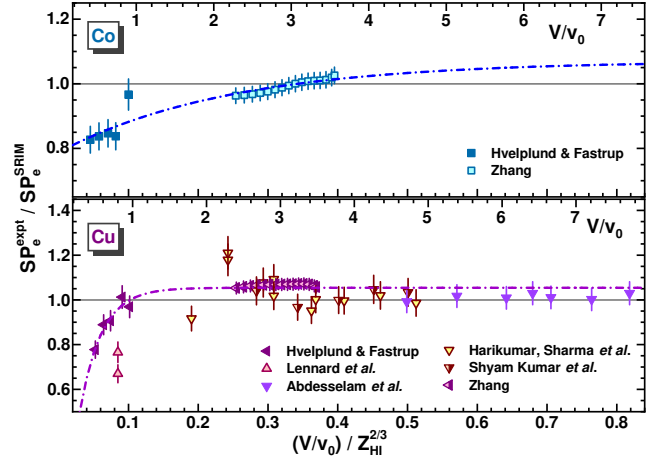}
\caption{\label{CoCuSPeSRIM} The same as in Figs.~\ref{ArKSPeSRIM}--\ref{MnFeSPeSRIM}, but for Co and Cu ions (upper and bottom panels, respectively). See the text for details.}
\end{center}
\end{figure}

\begin{figure}[!h]   
\begin{center}
\includegraphics[width=0.485\textwidth]{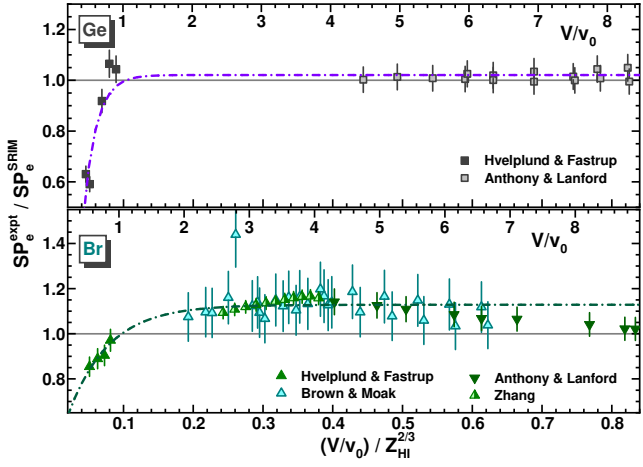}
\caption{\label{GeBrSPeSRIM} The same as in Figs.~\ref{ArKSPeSRIM}--\ref{CoCuSPeSRIM}, but for Ge and Br ions (upper and bottom panels, respectively). See the text for details.}
\end{center}
\end{figure}

\begin{figure}[!h]   
\begin{center}
\includegraphics[width=0.485\textwidth]{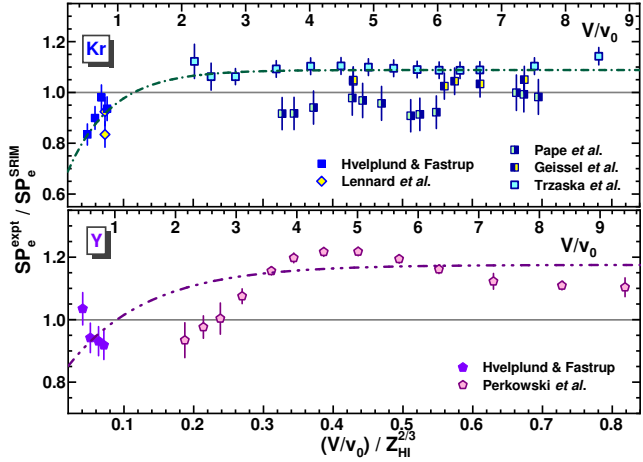}
\caption{\label{KrYSPeSRIM} The same as in Figs.~\ref{ArKSPeSRIM}--\ref{GeBrSPeSRIM}, but for Kr and Y ions (upper and bottom panels, respectively). See the text for details.}
\end{center}
\end{figure}

\begin{figure}[!h]   
\begin{center}
\includegraphics[width=0.485\textwidth]{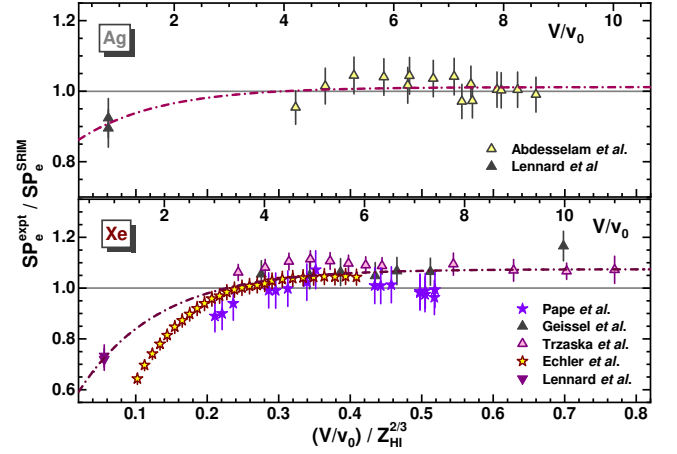}
\caption{\label{AgXeSPeSRIM} The same as in Figs.~\ref{ArKSPeSRIM}--\ref{KrYSPeSRIM}, but for Ag and Xe ions (upper and bottom panels, respectively). See the text for details.}
\end{center}
\end{figure}

\begin{figure}[!h]  
\begin{center}
\includegraphics[width=0.485\textwidth]{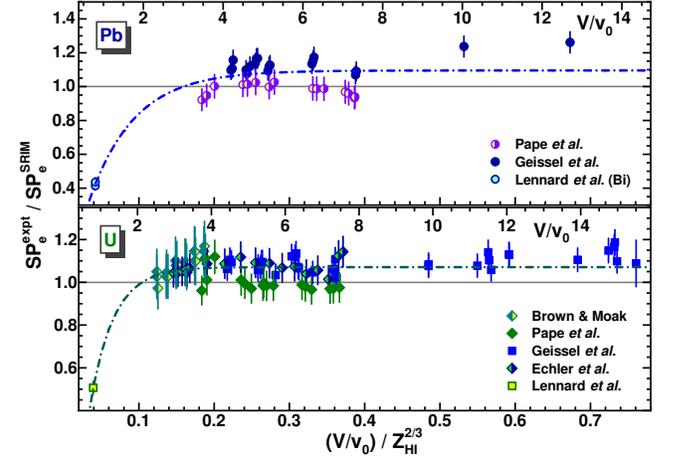}
\caption{\label{PbUSPeSRIM}The same as in Figs.~\ref{ArKSPeSRIM}--\ref{AgXeSPeSRIM}, but for Pb and U ions (upper and bottom panels, respectively). See the text for details.}
\end{center}
\end{figure}

In Figs.~\ref{ArKSPeSRIM}--\ref{PbUSPeSRIM}, comparisons of electronic SP values derived from experiments with those obtained from SRIM calculations are shown for HIs from Ar to U. Some details for $SP_{e}^{\rm expt}$ data used within this consideration should be mentioned, and they are discussed below. As with the $SP_{tot}^{\rm expt}$, the analysis favored the use of available original data (with the original errors).

The $SP_{e}^{\rm expt}/SP_{e}^{\rm SRIM}$ values for Ar to U ions were fitted using the weighted LSM procedure applied with the exponential correction function of $V_{r}$, similar to Eq.~(8):
\begin{equation}  \label{expSPefit}
 f_{cor}^{e}(V_{r}) = SP_{e}^{\rm expt}/SP_{e}^{\rm SRIM} = a_{e} - b_{e}\exp(-k_{e} V_{r}),
\end{equation}
where $a_{e}$, $b_{e}$, and $k_{e}$ are fitting parameters.

The ratios of the $SP_{e}^{\rm expt}$ and $SP_{e}^{\rm SRIM}$ values for Ar and K ions are shown in Fig.~\ref{ArKSPeSRIM}. The Ar and K $SP_{e}^{\rm expt}$ data \cite{Fastrup66,Lennard86,Pape78,Trzaska18,SchulzBrandt82,Sharma99} were taken from the works' respective tables, whereas the Ar data of Geissel \ea\ from the database \cite{IAEASP}. High velocity data \cite{Pape78,Sharma99} and those of Geissel \ea\ are the same as shown in Fig.~\ref{ArKSPtSRIM}, whereas the data \cite{Trzaska18} are limited to velocities $V_{r} > 0.3$. The limit corresponds to the value from which $SP_{n}^{\rm SRIM}$ gives a lower contribution than the SP data errors assigned in \cite{Trzaska18}. The Ar data \cite{Pape78} contradict similar ones \cite{Trzaska18} and those of Geissel \ea\ \cite{IAEASP}, which are in agreement, as shown in Fig.~\ref{ArKSPtSRIM}. The low velocity data \cite{Lennard86} also contradict the data \cite{Fastrup66,SchulzBrandt82}, as shown in Fig.~\ref{ArKSPeSRIM}. As a result, the data \cite{Lennard86,Pape78} were then removed from the fitting of $SP_{e}^{\rm expt}/SP_{e}^{\rm SRIM}$ ratio. Fitting parameter values thus obtained for the Ar and K data are listed in Table~\ref{abkeparam}.

Figure~\ref{CaScSPeSRIM} shows the Ca and Sc electronic SP data \cite{Fastrup66,Trzaska18,Sharma99,Perkowski06,ShyKum96} in comparison to SRIM calculations. As in Fig.~\ref{CaScSPtSRIM}, the fitting curve does not provide Ca data for the match with Eq.~(\ref{expSPefit}) even with omitting the data \cite{Sharma99} at $0.3 < V_{r} < 0.4$. Note that the $SP_{n}^{\rm SRIM}$ value at the lowest $V_{r}$ of the data \cite{Perkowski06} corresponds to 2\% of $SP_{tot}^{\rm SRIM}$, which is smaller than the SP data error (3.8\%) assigned in the work.  Thus, the data \cite{Perkowski06} could be related to $SP_{e}^{\rm expt}$ values in the whole range of $V_{r}$. It is implied that the electronic SP data \cite{Perkowski06} at relatively high velocities and those of \cite{Fastrup66} at relatively low velocities are reliable, as for the total SP considerations. The best Sc data fit was obtained with the constant SP ratio when the data \cite{Lennard86} were ignored. The latter correspond to noticeably lower electronic SP values than those obtained in \cite{Fastrup66}. The results of the Ca and Sc data fitting are listed in Table~\ref{abkeparam}.

Figure~\ref{TiCrSPeSRIM} shows the Ti and Cr electronic SP data \cite{Zhang2002,Hvelp68,Lennard86,Pape78,AnthLanf82,Abdess92,Sharma99,ShyKum96,Harikumar96} and those obtained by Geissel \ea\ in comparison to SRIM calculations. The Ti data at relatively high velocities ($V_{r} \gtrsim 0.3$) are in satisfactory agreement with each other (data \cite{AnthLanf82} and Geissel \ea\ were taken from the database \cite{IAEASP}). The Cr data \cite{Zhang2002} at the lowest velocities, for which the $SP_{n}^{\rm SRIM}$ values exceeded data errors (2.5\%), were excluded from fitting. The best data fit was obtained with the constant SP ratio when the value \cite{Lennard86} was disregarded. The last corresponds to the significantly lower electronic SP value than the one obtained in \cite{Fastrup66}. The results of the Ti and Cr data fitting are given in Table~\ref{abkeparam}.

Figure~\ref{MnFeSPeSRIM} compares the Mn and Fe electronic SP data \cite{Zhang2002,Hvelp68,Lennard86,ShyKum96,Sharma99,Harikumar96,Harikumar97} to SRIM calculations  (data \cite{Zhang2002} were taken from the database \cite{IAEASP}). The Mn data at $0.2 \lesssim V_{r} \lesssim 0.5$ \cite{ShyKum96,Sharma99} differ from those obtained later with minor errors \cite{Zhang2002}. The Mn datum at $0.8v_{0}$ \cite{Lennard86} does not agree with the data \cite{Hvelp68} at the same velocity and was excluded from fitting. The Mn and Fe data \cite{Zhang2002} at the lowest velocities, for which the $SP_{n}^{\rm SRIM}$ values exceeded data errors (2.5\%), were also excluded from fitting. The results of fitting are listed in Table~\ref{abkeparam}.

\begin{table}[!h]  
\caption{\label{abkeparam} Parameter values $a_{e}$, $b_{e}$, and $k_{e}$, as obtained for the Ar to U $SP_{e}^{\rm expt}/SP_{e}^{\rm SRIM}$ data fitted with Eq.~(\ref{expSPefit}). The ion symbols and ranges of Eq.~(\ref{expSPefit}) applicability for specified ions are listed in the first and last columns, respectively. The results of data fitting are also
shown in Figs.~\ref{ArKSPeSRIM}--\ref{PbUSPeSRIM}.
}
\begin{ruledtabular}
\begin{tabular}{lccrr}
Ion                                   &                    $a_{e}$       &                $b_{e}$       & \multicolumn{1}{c}{$k_{e}$} &$V_{r}$ range \\
\hline
  Ar\footnotemark[1] & 1.0490$\pm$0.0056 & 0.351$\pm$0.076      & 10.8$\pm$2.6   &  0.05--0.8    \\
   K\footnotemark[2]  & 0.9980$\pm$0.0095 & 0.725$\pm$0.474      & 35.5$\pm$12.0 &  0.05--0.6    \\
  Ca\footnotemark[3] & 1.0928$\pm$0.0279 & 0.236$\pm$0.043      &   2.9$\pm$1.3    &   0.06--0.9   \\
  Sc\footnotemark[2]  & 0.9989$\pm$0.0098 &                                          &                                 &   0.06--0.8   \\
  Ti                                    & 1.0580$\pm$0.0126  & 0.233$\pm$0.057      &   6.3$\pm$2.9    &  0.06--0.8    \\
  V                                     & 0.9852$\pm$0.0274  &                                          &                                 &    0.2--0.6    \\
  Cr\footnotemark[2]$^{,}$\footnotemark[4]  & 0.9725$\pm$0.0110 &                                           &                                 &    0.06--0.7  \\
 Mn\footnotemark[2]$^{,}$\footnotemark[4] & 1.0676$\pm$0.0526 & 0.231$\pm$0.053      &   4.3$\pm$3.0     &    0.06--0.7  \\
  Fe\footnotemark[4]  & 1.0278$\pm$0.0054 & 0.832$\pm$0.335      & 24.3$\pm$7.1     &   0.04--0.7  \\
  Co\footnotemark[4]  & 1.0707$\pm$0.0373 & 0.282$\pm$0.027     &    4.0$\pm$1.3     &   0.04--0.5  \\
  Cu\footnotemark[2]$^{,}$\footnotemark[4]  & 1.0542$\pm$0.0055 & 1.464$\pm$0.849     & 32.2$\pm$9.6      &   0.05--0.8  \\
  Ge                                    & 1.0207$\pm$0.0150 & 5.385$\pm$3.329     & 53.7$\pm$12.8   &   0.04--0.9  \\
  Br                                     & 1.1284$\pm$0.0081 & 0.697$\pm$0.256     & 17.3$\pm$6.1      &   0.04--0.9  \\
  Kr\footnotemark[5]   & 1.0883$\pm$0.0072 & 0.550$\pm$0.197     & 15.9$\pm$5.8      &   0.04--0.8  \\
   Y                                      & 1.1752$\pm$0.0239 &0.386$\pm$0.166      &   8.8$\pm$5.4      &   0.04--0.8  \\
  Ag                                    & 1.0116$\pm$0.0131 & 0.178$\pm$0.110     &    8.9$\pm$10.7   &   0.06--0.7  \\
    I                                     & 1.0786$\pm$0.0069 &                                          &                                   &    0.2--0.6    \\
  Xe                                    & 1.0744$\pm$0.0111 & 0.579$\pm$0.065     &    9.0$\pm$1.2      &   0.05--0.8   \\
  Pb                                    & 1.0953$\pm$0.0252 & 1.352$\pm$0.313     & 16.7$\pm$6.1      &   0.04--0.8   \\
   U                                     & 1.0709$\pm$0.0061 & 1.972$\pm$0.661     & 31.9$\pm$8.5      &   0.04--0.8   \\
\end{tabular}
\end{ruledtabular}
\footnotetext[1]{Without the data \cite{Lennard86,Pape78} (see the text).}
\footnotetext[2]{Without the data \cite{Lennard86}(see the text).}
\footnotetext[3]{Without the data \cite{Sharma99} at $0.3 < V_{r} < 0.4$ (see the text)}.
\footnotetext[4]{Without the data \cite{Zhang2002} at the lowest velocities (see the text).}
\footnotetext[5]{Without the data \cite{Pape78} (see the text).}
\end{table}

Figure~\ref{CoCuSPeSRIM} compares the Co and Cu electronic SP data \cite{Zhang2002,Hvelp68,Lennard86,Abdess92,ShyKum96,Harikumar96,Harikumar97} to SRIM calculations (the data \cite{Zhang2002} were taken from the database \cite{IAEASP}). The SP data \cite{Zhang2002} at the lowest velocities, at which the $SP_{n}^{\rm SRIM}$ values exceeded data errors (2.5\%), were excluded from fitting. The Cu data at $0.8v_{0}$ \cite{Lennard86} do not agree with the data \cite{Hvelp68} at the same velocity and were excluded from fitting. As a result, the fitting curves provide rather good matches for the Co and Cu data \cite{Zhang2002} with those obtained at low velocities \cite{Hvelp68}. At the same time, these curves vary markedly. The results of fitting are listed in Table~\ref{abkeparam}.

Figure~\ref{GeBrSPeSRIM} compares the Ge and Br electronic SP data \cite{Zhang2002,Hvelp68,BrownMoak72,AnthLanf82} to SRIM calculations (the data \cite{Zhang2002} were taken from the database \cite{IAEASP}). The Br data at $V_{r} \lesssim 0.6$ \cite{Zhang2002,BrownMoak72,AnthLanf82} are reasonably consistent. As in previous cases, the Br data \cite{Zhang2002} at the lowest velocities, for which the $SP_{n}^{\rm SRIM}$ values exceeded data errors (2.5\%), were excluded from fitting. As in the case of the Co and Cu ratios, fitting curves obtained for the Ge and Br ratios vary markedly. The results of fitting are listed in Table~\ref{abkeparam}.

Figure~\ref{KrYSPeSRIM} shows the Kr and Y electronic SP data \cite{Hvelp68,Lennard86,Pape78,Trzaska18,Perkowski09} and those obtained by Geissel \ea\ (taken from the database \cite{IAEASP}) in comparison to SRIM calculations. As for the Ar data, the Kr data \cite{Pape78} at $V_{r}\gtrsim 0.3$ lie noticeably below the data of Geissel \ea\ and \cite{Trzaska18}, which are in satisfactory agreement with each other. Note that the errors for the Kr data points at the lowest velocities \cite{Trzaska18} exceed the respective $SP_{n}^{\rm SRIM}$ values. The data \cite{Trzaska18}, together with the data of Geissel \ea\ and the low velocity data \cite{Hvelp68,Lennard86}, are well fitted with Eq.~(\ref{expSPefit}), as shown in the figure. In the fitting procedure, the Y data \cite{Perkowski09} were restricted to the velocities for which the $SP_{n}^{\rm SRIM}$ values did not exceed data errors. These data do not match to the low velocity data \cite{Hvelp68} [$\chi^{2}_{r} = 8.50$ for fitting with Eq.~(\ref{expSPefit})]. Implying the reliability of the data \cite{Perkowski09}, a bad data fit could be explained by the oversimplified fitting model, which does not allow us to describe the data in a wide range of $V_{r}$.

Figure~\ref{AgXeSPeSRIM} shows the electronic Ag and Xe SP data \cite{Lennard86,BrownMoak72,Pape78,AnthLanf82,Abdess92,Trzaska18,Echler17}, and those obtained by Geissel \ea\ (taken from the database \cite{IAEASP}) in comparison to SRIM calculations. Because nuclear stopping accounts for only 0.8\% of total stopping for Ag at the lowest velocity, the Ag data \cite{Abdess92} were assigned to the $SP_{e}^{\rm expt}$ values. This value is much lower than the 5\% total uncertainty assigned to the data. The Xe data \cite{Trzaska18} at the lowest velocities, for which the $SP_{n}^{\rm SRIM}$ values exceeded the data errors, were excluded from fitting. The original $SP_{e}^{\rm expt}$ data for $V_{r} \gtrsim 0.2$ \cite{Echler17} were only used for Xe data fitting. At these velocities, $SP_{n}^{\rm SRIM}$ values contribute less than 3\% of $SP_{tot}^{\rm SRIM}$. That is less than the respective data errors assigned in \cite{Echler17}. TRIM simulations were used for the estimates of the $SP_{n}$ values in \cite{Echler17} with subsequent subtraction from the measured $SP_{tot}^{\rm expt}$. The low velocity data at $V_{r} \lesssim 0.2$ \cite{Echler17}, shown in the figure, disagree with the data \cite{Lennard86} due to the overestimate of nuclear stopping in TRIM simulations. At the same time, all the Xe data at $V_{r} > 0.2$ satisfactory agree with each other. The results of fitting with Eq.~(\ref{expSPefit}) are listed in Table~\ref{abkeparam}.

The iodine stopping power data \cite{BrownMoak72,AnthLanf82,Abdess92} at $V_{r} > 0.2$ (considered in Sec.~\ref{SPtotSRIM}) are in rather good agreement with each other, whereas at $V_{r} < 0.2$ the data \cite{BrownMoak72,AnthLanf82} are varied (see Fig.~\ref{AgISPtSRIM}). At $V_{r} > 0.2$, the $SP_{n}^{\rm SRIM}$ values are less than 3\% of $SP_{tot}^{\rm SRIM}$ and less than the data errors (5--10\%). Thus, the electronic SP ratios at $V_{r} > 0.2$ for I ions was the same as the one obtained by the constant fit to the iodine $SP_{tot}^{\rm expt}/SP_{tot}^{\rm SRIM}$ ratios (see Table~\ref{abkeparam}).

Figure~\ref{PbUSPeSRIM} shows the electronic Pb and U SP data \cite{Lennard86,BrownMoak72,Pape78,Echler12}, and those obtained by Geissel \ea\ (taken from the database \cite{IAEASP}) in comparison to SRIM calculations. Though the Pb data of Geissel \ea\ and \cite{Pape78} are in some disagreement with each other, they were fitted with Eq.~(11) altogether, as for the total SP ratios. As in previous cases, the data of Geissel \ea\ at the lowest velocities, for which the $SP_{n}^{\rm SRIM}$ values exceeded 5\% data errors, were excluded from the fitting procedure. That is the same for the U data of the same authors. The lasts, together with others \cite{BrownMoak72,Echler12}, seem to be in satisfactory agreement with each other.  The U data \cite{Echler12} at the lowest velocities, for which the $SP_{n}^{\rm SRIM}$ values exceeded 6\% errors, were excluded from fitting. The Pb and U data were supplemented by the Bi and U data obtained at $0.8 v_{0}$ \cite{Lennard86}. In doing so, the possible distinction in the SP values for Pb and Bi was neglected. The results of fitting are listed in Table~\ref{abkeparam}.

\begin{figure}[!h]  
\begin{center}
\includegraphics[width=0.4\textwidth]{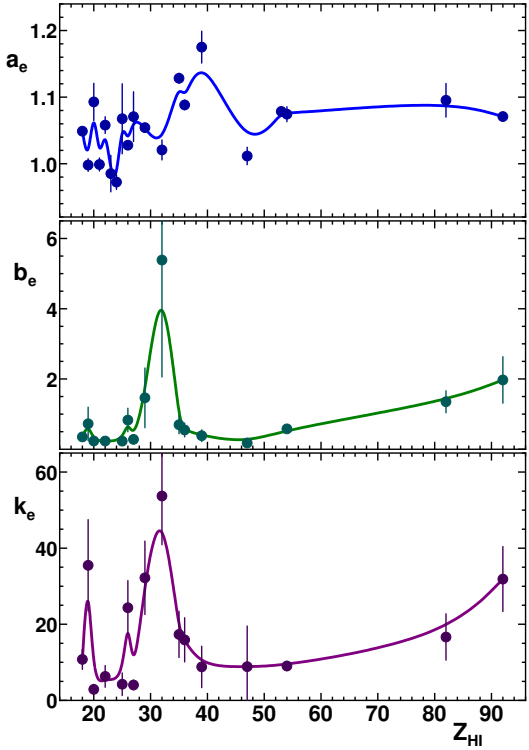}
\caption{\label{abkepar}The same as in Fig.~\ref{abkpar}, but for fitting parameters $a_{e}$, $b_{e}$, and $k_{e}$ listed in Table~\ref{abkeparam}.}
\end{center}
\end{figure}

In Fig.~\ref{abkepar}, the fitting parameters listed in Table~\ref{abkeparam} are shown as a function of the HI atomic number. Amplitude $a_{e}$, corresponding to the $SP_{e}^{\rm expt}/SP_{e}^{\rm SRIM}$ ratio at high velocities, oscillates in a sporadic way within 0.8--1.2, i.e., similarly to the $SP_{tot}^{\rm expt}/SP_{tot}^{\rm SRIM}$ ratio. The exponent parameter values ($b_{e}$ and $k_{e}$), which determine decreasing the $SP_{e}^{\rm expt} / SP_{e}^{\rm SRIM}$ at low velocities, correlate with each other to a certain extent in the region of $24 \lesssim Z_{\rm HI} \lesssim 40$ and probably for higher $Z_{\rm HI}$. These correlations are similar to the $SP_{tot}^{\rm expt}/SP_{tot}^{\rm SRIM}$ ones (see Fig.~\ref{abkpar}).

\begin{figure}[!h]  
\begin{center}
\includegraphics[width=0.485\textwidth]{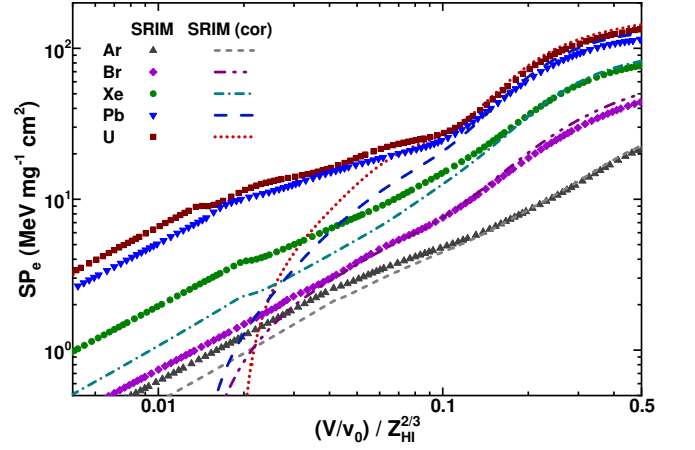}
\caption{\label{SPecorr}The $SP_{e}$ SRIM values corrected with Eq.~(\ref{expSPefit}) (lines) are shown in comparison with the original ones \cite{SRIM} (symbols) for Ar, Br, Xe, Pb, and U ions.}
\end{center}
\end{figure}

In Fig.~\ref{SPecorr}, some examples of the corrected $SP_{e}$ SRIM values are shown in comparison to the original ones \cite{SRIM}. As seen, the corrections become significant for Xe and heavier ions at velocities $(V/v_{0})/Z_{\rm HI}^{2/3} \lesssim 0.1 $.

Now, with the correction functions for $SP_{tot}^{\rm SRIM}$ and $SP_{e}^{\rm SRIM}$, Eq.~(\ref{SPnexpt}) can be rewritten as
\begin{equation}
  SP_{n}^{\rm expt} = f_{cor}(V_{r}) SP_{tot}^{\rm SRIM} - f_{cor}^{e}(V_{r})SP_{e}^{\rm SRIM},          \label{SPnex}
\end{equation}
and thus the nuclear stopping powers could be empirically estimated for specified HIs. Respective SP values are  determined by the parameter values of the $f_{cor}(V_{r})$ and $f_{cor}^{e}(V_{r})$ functions listed in Tables~\ref{abkparam} and \ref{abkeparam}.

\subsection{\label{SPnest}Nuclear stopping power estimates}

\begin{figure}[!h]  
\begin{center}
\includegraphics[width=0.485\textwidth]{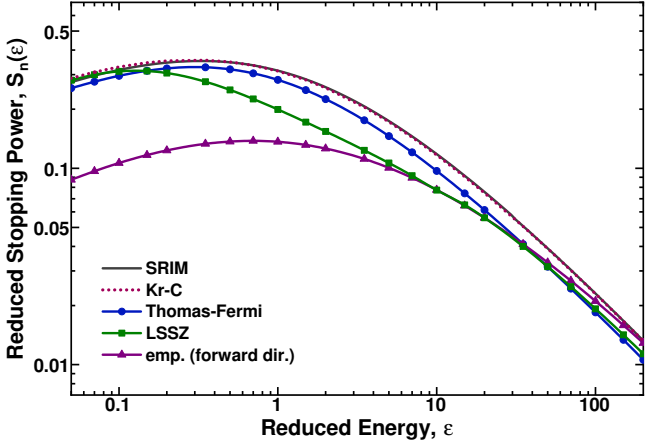}
\caption{\label{Snapprox} Some approximations for reduced nuclear stopping power $S_{n}(\varepsilon)$ are shown: the one used in SRIM calculations (solid line), and those considered in Sec.~\ref{LSScomp} and this section. The lasts are designated as ``Thomas-Fermi'', ``LSSZ'' \cite{Ziegler77}, ``emp. (forward dir.)'' \cite{Garnir80} (solid lines with symbols) and ``Kr-C'' \cite{Wilson77} (dotted line). See the text for details.}
\end{center}
\end{figure}

For the orientation, different nuclear SP approximations mentioned in Sec.~\ref{LSScomp} are shown in Fig.~\ref{Snapprox} in the common form of reduced $S_{n}(\varepsilon)$ functions. These functions are compared with the ``universal'' nuclear stopping power used in SRIM calculations:
\begin{eqnarray}\label{SnSRIM}
  S_{n}^{\rm SRIM}(\varepsilon) & = & \frac{0.5\ln(1 + 1.1383 \varepsilon)}{\varepsilon + 0.01321\varepsilon^{0.21226} + 0.19593\varepsilon^{0.5}}, \varepsilon \leqslant 30;  \nonumber \\
  S_{n}^{\rm SRIM}(\varepsilon) & = & 0.5 \ln(\varepsilon)/\varepsilon, \phantom{3.5 \ln(\varepsilon)/\varepsilon\ln(\varepsilon)} \varepsilon > 30;
\end{eqnarray}
where the reduced energy is determined as
\begin{equation}\label{epsSRIM}
  \varepsilon = 32.53A_{t}E/[(A_{\rm HI}+A_{t})Z_{\rm HI}Z_{t}(Z_{\rm HI}^{0.23}+Z_{t}^{0.23})].
\end{equation}

In addition, a pure empirical formula \cite{Garnir80}, corresponding to measurements made in a forward direction along the beam axis within a narrow acceptance angle (as mentioned in Sec.~\ref{Intromotiv}):
\begin{equation}\label{Snemp}
  S_{n}^{\rm emp} = 0.75 \frac{\ln(0.78 \varepsilon^{0.5} + 1)}{(0.78 \varepsilon^{0.5} + 1)^{2}},
\end{equation}
where $\varepsilon$ is determined by Eq.~(\ref{epsE}), is also shown in Fig.~\ref{Snapprox}. The approximation expressed by Eq.~(\ref{SnWHB}) (considered as ``the best'' one in \cite{Wilson77}) is also added to the figure. It uses $\varepsilon$ values corresponding to  a screening length
\begin{equation}\label{aKrC}
  a_{scr} = 0.8853 a_{0} / (Z_{\rm HI}^{1/2}+Z_{t}^{1/2})^{2/3},
\end{equation}
where $a_{0}$ is the Bohr radius, whereas $\varepsilon$ values determined by Eq.~(\ref{epsE}) use another screening length definition \cite{Sigmund04}.

As one can see in the figure, all approximations give close $S_{n}$ values at $\varepsilon \gtrsim 10$, whereas at $\varepsilon \lesssim 10$ they are varied, and the difference between $S_{n}^{\rm emp}$ reaches a factor of $\sim$3 at $\varepsilon = 0.05$. In further consideration, $S_{n}(\varepsilon)$ dependencies derived with the transformation of $SP_{n}^{\rm expt}$ obtained with Eq.~(\ref{SPnex}) into reduced $S_{n}$ values are compared with the SRIM and empirical approximations given by Eqs.~(\ref{SnSRIM}) and (\ref{Snemp}), respectively.

\begin{figure}[!h]  
\begin{center}
\includegraphics[width=0.485\textwidth]{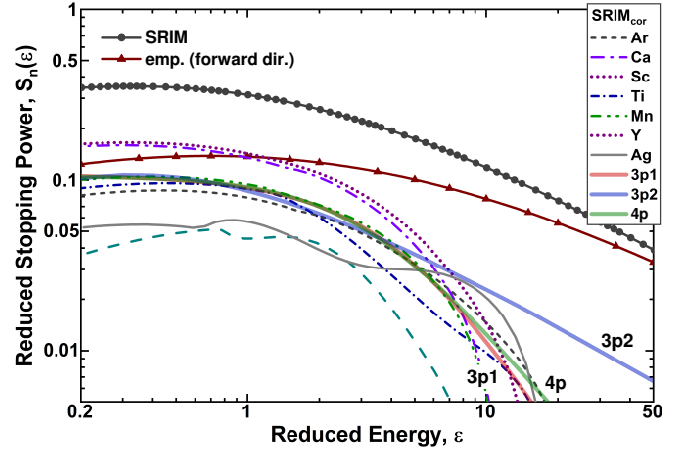}
\caption{\label{SnArAg} The $S_{n}(\varepsilon)$ dependencies obtained with Eq.~(\ref{SPnex}) for Ar, Ca, Sc, Ti, Mn, Y, and Ag are shown by respective intermittent lines in comparison to SRIM and empirical approximations \cite{SRIM,Garnir80} corresponding to Eqs.~(\ref{SnSRIM}) and (\ref{Snemp}) (solid lines with circles and triangles, respectively). The $S_{n}(\varepsilon)$ dependencies for the Ar--Ag set, as obtained with Eqs.~(\ref{ABC1fit}) -- (\ref{ABCDfit}), are shown by thick solid (denoted as 3p1, 3p2, and 4p).}
\end{center}
\end{figure}

\begin{figure}[!h]  
\begin{center}
\includegraphics[width=0.485\textwidth]{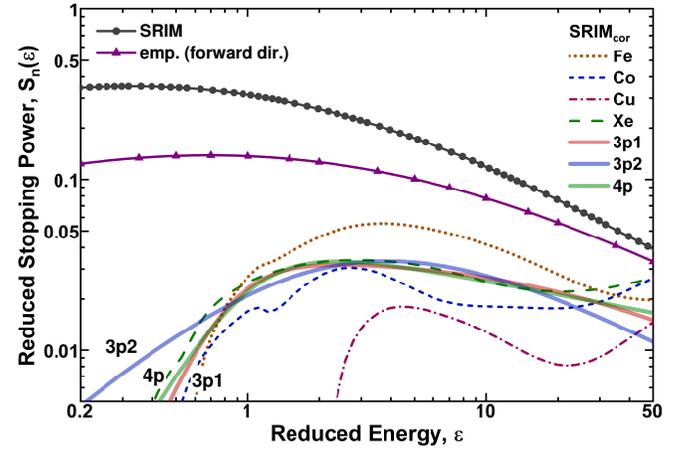}
\caption{\label{SnFeXe} The same as in Fig.~\ref{SnArAg}, but for Fe, Co, Cu, and Xe.}
\end{center}
\end{figure}

\begin{figure}[!h]  
\begin{center}
\includegraphics[width=0.485\textwidth]{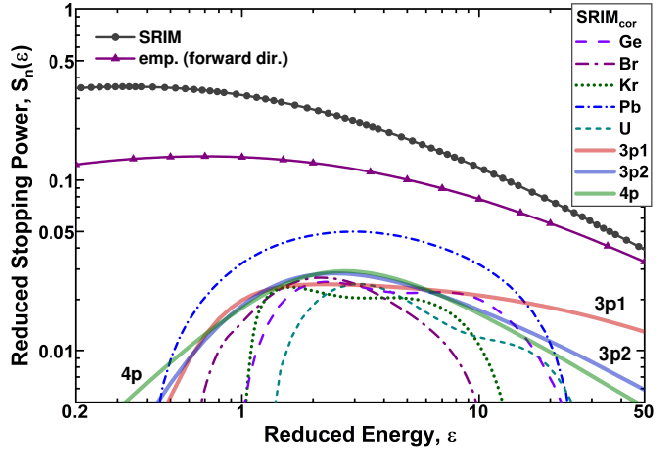}
\caption{\label{SnGeU} The same as in Figs.~\ref{SnArAg} and \ref{SnFeXe}, but for Ge, Br, Kr, Pb, and U.}
\end{center}
\end{figure}

Three different approximations, corresponding to three groups of HIs, arise as a result of the application of Eq.~(\ref{SPnex}) and subsequent data conversion to the reduced $S_{n}(\varepsilon)$ functions. These seem to be quite unexpected results, which are shown in Figs.~\ref{SnArAg}--\ref{SnGeU}.

The $S_{n}(\varepsilon)$ dependencies for Ar, Ca, Sc, Ti, Mn, Y and Ag ions are shown in Fig.~\ref{SnArAg}. These dependencies have a similar behavior within a factor of $\sim$2. At $\varepsilon < 2$, the $S_{n}$ values for Ar, Ca, Sc, Ti, and Y are close to those given by Eq.~(\ref{Snemp}). All the dependencies show a steep fall in the $S_{n}$ values at $\varepsilon > 2$, as compared with SRIM and empirical approximations [Eqs.~(\ref{SnSRIM}) and (\ref{Snemp}), respectively]. In view of uncertainties in the $SP_{n}^{\rm expt}$ estimates, this data set and two others could be approximated by the ``universal'' expressions \cite{Wilson77} fitted with respective parameters:
\begin{eqnarray}
  S_{n} &=& [0.5 \ln(1 + A \varepsilon)] / (\varepsilon + B \varepsilon^C), \label{ABC1fit}  \\
  S_{n} &=& [A \ln(B \varepsilon)] / [B \varepsilon - 1 / (B \varepsilon)^C],  \label{ABC2fit}  \\
  S_{n} &=& 1/(4 + A / \varepsilon^B + C /  \varepsilon^D),  \label{ABCDfit}
\end{eqnarray}
where $A$, $B$, $C$, and $D$ are fitting parameters. The unweighted LSM for fitting was applied to such $S_{n}(\varepsilon)$ dependencies relating to the specific set of HIs. The $S_{n}(\varepsilon)$ changes were limited by ranges of $0.2 \lesssim \varepsilon \lesssim 50$ and $S_{n} \gtrsim 0.005$. The parameter values thus obtained for the Ar--Ag set are listed in Table~\ref{SnABCDfit}. The 3p1 and 4p expressions [Eqs.~(\ref{ABC1fit}) and (\ref{ABCDfit}), respectively] correspond to the best fits, as seen in Fig.~\ref{SnArAg} for this data set.

In Fig.~\ref{SnFeXe}, the $S_{n}(\varepsilon)$ dependencies for Fe, Co, Cu, and Xe ions are shown. They vary within a factor of $\sim$3 at $\varepsilon > 3$ and show a steep fall in the $S_{n}$ values at $\varepsilon < 3$, as compared to the SRIM and empirical approximations. The best fits correspond to the 3p1 and 4p expressions, as in the previous case. Fitted parameter values for this Fe--Xe set are listed in Table~\ref{SnABCDfit}. The unexpected steep fall in the $S_{n}$ values at $\varepsilon < 3$ implies that nuclear stopping plays a minor role for this HI set at low energies. Such behavior of $S_{n}(\varepsilon)$ is in contrast to the one observed for the previous Ar--Ag set, which seems to be reasonable at low energies. At the same time, $S_{n}(\varepsilon)$ approaches the SRIM and empirical approximations at $\varepsilon \gg 3$, giving lower values than those obtained by these approximations by a factor of 2--4, unlike the previous case.

\begin{table}[!h]  
\caption{\label{SnABCDfit} The $A$, $B$, $C$, and $D$ parameter values are listed as the result of fitting of $S_{n}(\varepsilon)$ with Eqs.~(\ref{ABC1fit})--(\ref{ABCDfit}) applied to the specified ion sets. The dependencies were combined into three sets (indicated in the first column) corresponding to the similarity in the $S_{n}(\varepsilon)$ behavior. The results of data fitting are also shown in Figs.~\ref{SnArAg}--\ref{SnGeU}.
}
\begin{ruledtabular}
\begin{tabular}{cccccc}
HI set    &     Eq.                &               $A$               &                $B$          &                  $C$             &           $D$                  \\
\hline
  Ar--Ag & \ref{ABC1fit} & 0.217$\pm$0.008 &  0.09$\pm$0.04  & 2.66$\pm$0.31       &                                      \\
                & \ref{ABC2fit} & 0.224$\pm$0.035 &   3.50$\pm$1.08 & 1.10$\pm$0.35       &                                      \\
                & \ref{ABCDfit} &   1.53$\pm$0.46    & -1.66$\pm$0.22  & 5.47$\pm$0.35      & 0\footnotemark[1] \\
  Fe--Xe & \ref{ABC1fit}  & 0.070$\pm$0.003 &   0.47$\pm$0.16  & -2.37$\pm$0.88    &                                      \\
               & \ref{ABC2fit} & 0.079$\pm$0.006 &  0.43$\pm$0.09    & 1.51$\pm$0.25      &                                      \\
               & \ref{ABCDfit} &  18.2$\pm$2.4       & -0.29$\pm$0.05    &  22.9$\pm$5.2       &   2.43$\pm$0.60     \\
  Ge--U  & \ref{ABC1fit} & 0.053$\pm$0.003 &   0.30$\pm$0.17    & -2.69$\pm$1.53    &                                      \\
               & \ref{ABC2fit} & 0.076$\pm$0.004 &   1.00$\pm$0.15    &  3.12$\pm$0.90     &                                      \\
               & \ref{ABCDfit} &  44.2$\pm$6.2       &   1.31$\pm$0.40    &  7.91$\pm$3.95     &   -0.84$\pm$0.26   \\
\end{tabular}
\end{ruledtabular}
\footnotetext[1]{Fixed value, corresponding to the best fit.}
\end{table}

The $S_{n}(\varepsilon)$ dependencies for Ge, Br, Kr, Pb, and U ions are shown in Fig.~\ref{SnGeU}. In contrast to previous cases, nuclear stopping is only manifested in the region of $0.5 \lesssim \varepsilon \lesssim 15$. The $S_{n}(\varepsilon)$ approximations fitted to these dependencies have a maximum at $\varepsilon \simeq 3$. The maximum value is lower than the values given by the SRIM and empirical approximations at this energy by a factor of 4--6. This behavior of $S_{n}(\varepsilon)$ seems to be an unexpected one. Unlike in previous cases, the $S_{n}(\varepsilon)$ dependencies obtained with Eqs.~(\ref{ABC1fit})--(\ref{ABCDfit})  do not provide acceptable approximations at low and high energies.

\section{\label{discus}Discussion}

A quantitative comparison of the SP data obtained at $0.8v_{0}$ to similar ones obtained with SRIM/TRIM model calculations/simulations \cite{SRIM} showed the disadvantages of the latter in reproducing both the total and electronic SP values for HIs. For ions heavier than Xe, the model overestimates SP by a factor of 2.5 as compared to the data \cite{Lennard86}, whereas for the lighter ions the disagreements amount within +5 to -20\% of the data \cite{Fastrup66,Hvelp68} and within -10 to -30\% for the data \cite{Lennard86} (see Figs.~\ref{SPeSRIMcomp} and \ref{dEdXSPecomp}). The $\sim$20\% difference between the data \cite{Fastrup66,Hvelp68} and \cite{Lennard86} could be explained by the difference in experimental conditions (difference in the detection angle in forward direction, relative to the beam axis for HIs escaping a target). One more explanation is the different options for accounting for the nuclear (elastic) SP in a subsequent subtraction of this value from the measured one to obtain the sought electronic SP component.

In describing other results of the experiments \cite{Fastrup66,Hvelp68,Lennard86}, TRIM simulations showed a remarkable overestimate of the energy distribution width for Ar ions passing through a thin target ($\sim$7 $\mu$g/cm$^{2}$) in comparison to the value obtained in the experiment \cite{Fastrup66} (see Fig.~\ref{ArTRIM}).  That was in contrast to a similar comparison for Xe ions passing through a thick target ($\sim$30 $\mu$g/cm$^{2}$), for which the measured \cite{Lennard86} and simulated widths were about the same (see Fig.~\ref{XeTRIM}). These comparisons were conducted for HIs escaping the targets in a forward direction (the same conditions as in measurements). Simulations of the measurements at all angles, showed that one may expect 8.3 and 4.6\% increase in total SP values for Ar and Xe ions, respectively (see Figs.~\ref{ArTRIM} and \ref{XeTRIM}).

Some differences in total SP values for Cu ions for a thin ($\sim$5 $\mu$g/cm$^{2}$) and a relatively thick target ($\sim$30 $\mu$g/cm$^{2}$) were revealed in comparison of TRIM simulations to the experiments \cite{Hvelp68,Lennard86}. Thus, overestimates in simulations for the thin target corresponded to about 6.5\% and 48\% as compared to the experimental data \cite{Hvelp68} and \cite{Lennard86}, respectively. Simulations also showed SP independence of the output angle in a forward direction for HIs escaping the target, at least at $ 0.15^{\circ} \lesssim \theta_{\rm out} \lesssim 1.5^{\circ}$. For the relatively thick target, however, TRIM overestimated the measurements \cite{Lennard86} by 28\%.

The result of the analysis, as a whole, allowed us to state that SRIM/TRIM calculations/simulations remarkably overestimate carbon total stopping powers at the HI velocity of $0.8 v_{0}$ \cite{Lennard86}. At the same time, the electronic SP data for C to Y ions \cite{Fastrup66,Hvelp68} are in satisfactory agreement with calculations/simulations.

The data \cite{Fastrup66,Hvelp68,Lennard86} considered within the LSS approach \cite{LSS63}, as an alternative to SRIM/TRIM calculations/simulations, were not reproduced using different approximations to nuclear stopping. Although the electronic SP data could be described ``on average,'' the $Z_{\rm HI}$ oscillations observed in the experiments \cite{Fastrup66,Hvelp68,Lennard86} were not reproduced in the calculations. The addition of the calculated nuclear SPs \cite{Ziegler77} led to a deterioration of agreement with the experimental total SP data. Exceeding the data implies overestimates in the nuclear SP values.

Further analysis of the total and electronic SP data revealed that reduced nuclear stopping powers $S_{n}(\varepsilon)$ for HIs from Ar to U at $0.2 \leqslant \varepsilon \leqslant 50$ were generally lower than those predicted by any approximation \cite{SRIM,Garnir80,LSS63,Ziegler77,Wilson77,LNS68}. This reduction depends on the reduced energy (velocity) of HI and, unexpectedly, on the HI atomic number. The last dependence appeared in similarity of $S_{n}(\varepsilon)$ variations for the specified HI and enabled their grouping into clusters (sets).

Thus, for the conditionally first set of Ar, Ca, Sc, Ti, Mn, Y, and Ag, the $S_{n}(\varepsilon)$ dependencies for Ar, Ca, Sc, Ti, and Y, at $\varepsilon < 2$, approach the empirical formula \cite{Garnir80} derived from the data obtained in forward direction relative to the beam axis. All the dependencies show a steep fall in the $S_{n}$ values at $\varepsilon > 2$, as compared to the SRIM and empirical approximations \cite{SRIM,Garnir80}.

The $S_{n}(\varepsilon)$ values in the conditional second set of Fe, Co, Cu, and Xe differ from each other  by a factor of $\sim$3 at $\varepsilon > 3$ and show a steep fall in the $S_{n}$ values at $\varepsilon < 3$, as compared to the SRIM and empirical approximations. Such unexpected behavior implies a minor role for nuclear stopping for this set of HIs at low energies. This behavior of $S_{n}(\varepsilon)$ differs from the one observed in the first set that appears to be reasonable at low energies. However, unlike the previous set, the fitted $S_{n}(\varepsilon)$ dependencies approach to the SRIM and empirical approximations at $\varepsilon \gg 3$, while remaining lower than the values given by these approximations by a factor of 2--4.

For the conditionally third set of Ge, Br, Kr, Pb, and U, unlike in both previous cases, nuclear stopping only appears in the energy range of $0.5 \lesssim \varepsilon \lesssim 15$. The maximum values for the fitted $S_{n}(\varepsilon)$ dependencies are below the values given by the SRIM and empirical approximations by a factor of 4--6 at respective energies. Such behavior of $S_{n}(\varepsilon)$ turned out to be unexpected as well.

Thus, one can repeat the statement that the nuclear (collisional) carbon SP is remarkably lower than those predicted by any of the ``universal'' approximations \cite{SRIM,Garnir80,LSS63,Ziegler77,Wilson77,LNS68}, at least for HIs under consideration in the energy range of $0.2 \leqslant \varepsilon \leqslant 50$. This lowering depends on  $\varepsilon$ and, unexpectedly, on the HI atomic number.

In search of the causes of such nuclear SP behavior, it should be noted that nuclear stopping is determined by interatomic potential $U(r)$. These potentials were determined for 14 atomic systems using approximations to the results obtained by the free-electron method \cite{Wilson77}. These approximations were expressed as
\begin{equation}\label{ionpot}
  U(r/a_{scr}) = (Z_{\rm HI} Z_{t} e^{2}/r) \sum_{i=1}^{3}C_{i} \exp(-b_{i}r/a_{scr}),
\end{equation}
where $r$ is the interatomic separation, $e$ is the electronic charge, $C_{i}$ and $b_{i}$ are fitting constants, and $a_{scr}$ is a screening radius (length) according to Eq.~(\ref{aKrC}).

In TRIM simulations, interatomic interactions are determined by the ``Ziegler-Biersack-Littmark (ZBL) Universal Screening Potential'' \cite{SRIMbook} (citation according to Ref.~\cite{Paul13AIP}). The ZBL potential is similar to Eq.~(\ref{ionpot}), but it uses 8 fitting parameters ($i = 4$) and ``is based on calculated solid-state interatomic potentials of 522 randomly chosen pairs of atoms over the range of 1--82 for $Z_{\rm HI}$ and $Z_{t}$'' \cite{Paul13AIP}. A screening length is defined as \cite{Paul13AIP,Sigmund04}
\begin{equation}\label{aZBL}
  a_{scr} = 0.885 a_{0} / (Z_{\rm HI}^{0.23}+Z_{t}^{0.23}).
\end{equation}

One can speculate about reproducing the $S_{n}(\varepsilon)$ variations obtained here in calculations using the interatomic potential(s) according to Eq.~(\ref{ionpot}). Thus, one may mean specific values of the $C_{i}$, $b_{i}$ and $a_{scr}$ parameters corresponding to a specific potential for two colliding atoms. In this regard, the energy dependence of the ion charge inside the matter will determine the $a_{scr}(\varepsilon)$ dependence.

The interatomic potentials corresponding to  Eq.~(\ref{ionpot}) do not take into account the atomic shell structure of HIs moving inside the matter. This structure may affect ion charge states, making them different from those determined by $a_{scr}$ or SP values. The average values of charges $\overline{q}_{\rm HI}$ were estimated at $\varepsilon = 3$ (the value corresponding to the $S_{n}$ maxima for the Fe--Xe and Ge--U groups, as shown in Figs.~\ref{SnFeXe} and \ref{SnGeU}, respectively) in attempts to associate the values with the specific atomic (sub-) shell ionization. It was done with the relation used for the SP calculations within the effective (internal) charge concept \cite{SRIM,BrownMoak72,Ziegler77,AnthLanf82,Abdess92,Saga2015,Ziegler80,Barb2010_2}:
\begin{equation}\label{qinHIs}
  \overline{q}_{\rm HI} =  \overline{q}_{\rm H} (SP_{\rm HI} / SP_{\rm H})^{1/2},
\end{equation}
where index H refers to hydrogen. With the use of SRIM stopping powers and $\overline{q}_{\rm H}$ calculations \cite{Ziegler80}, we thus obtained $\overline{q}_{\rm HI} = 1.5-1.6$ for Ar to Co ions. The value implies the main charge states $q^{i}_{\rm HI} = +1, +2$ that, in turn, implies the potentials taking into account the filled Ar shell and outer electrons of the $3d$ sub-shell for Sc and heavier atoms. With this assumption $S_{n}(\varepsilon)$ could behave similarly for HIs ranging from Sc to Co. It is not the case, since the averaged $S_{n}(\varepsilon)$ function for Sc--Mn (Ar--Ag set) differs from the one for Fe and Co (Fe--Xe set), as shown in Figs.~\ref{SnArAg} and \ref{SnFeXe}. One may think that other aspects of interatomic interaction play an important role.

Quantitative estimates of the nuclear (collisional) stopping power component of carbon for heavy ions were considered using available data for 16 low-energy projectiles from Ar to U ($0.2 \leqslant \varepsilon \leqslant 50$). These data correspond to 21\% of all projectiles, for which stopping powers could be studied. The lack of stopping power data for most heavy ions could be filled by the consideration of their ranges \cite{Powers68,Grande88}, and the ranges of evaporation residues (ERs) produced in the nuclear fusion reactions induced by HIs \cite{SagaRange19}. A remarkable excess of the ranges for Tb, Dy, Po, At, Rn, and Ac ERs stopped in Al over those predicted by SRIM/TRIM in calculations/simulations was observed in the last study at energies $E < 0.1$ MeV/nucleon. These and similar artificial low-energy radioactive atoms are inaccessible as projectiles used in experiments on stopping power  measurements, whereas they can be produced in sufficient amounts in nuclear fusion reactions, allowing to extend studies for the most heavy atoms.


\begin{thebibliography}{99}

\bibitem{SRIM}  
 J.F.~Ziegler, \url{http://srim.org/}.

\bibitem{IAEASP}  
 \url{https://www-nds.iaea.org/stopping/};
 C.C.~Montanari and P.~Dimitriou,
 Nucl. Instrum. Methods Phys. Res. B \textbf{408}, 50 (2017).

\bibitem{Paul2003}  
 H.~Paul and A.~Schinner,
 Nucl. Instrum. Methods Phys. Res. B \textbf{209}, 252 (2003).

\bibitem{Paul2010}  
 H.~Paul,
 Nucl. Instrum. Methods Phys. Res. B \textbf{268}, 3421 (2010).

\bibitem{Paul2013}  
 H.~Paul and D.~S\'{a}nchez-Parscerisa,
 Nucl. Instrum. Methods Phys. Res. B \textbf{312}, 110 (2013).

\bibitem{Paul13AIP}  
 H.~Paul,
 AIP Conf. Proc. \textbf{1525}, 309 (2013).

\bibitem{Zhang2002}  
 Y.~Zhang,
 Nucl. Instrum. Methods Phys. Res. B \textbf{196}, 1 (2002).

\bibitem{Barb2010}  
 M.~Barbui, D.~Fabris, M.~Lunardon, S.~Moretto, G.~Nebbia, S.~Pesente, G.~Viesti, M.~Cinausero, G.~Prete, V.~Rizzi, K.~Hagel, S.~Kowalski, J.B.~Natowitz, L.~Qin, R.~Wada, and Z.~Chen,
 Nucl. Instrum. Methods Phys. Res. B \textbf{268}, 20 (2010).

\bibitem{Fastrup66}  
 B.~Fastrup, P.~Hvelplund, and C.A.~Sautter,
 K. Dan. Vidensk. Selsk. Mat. Fys. Medd. \textbf{35}, no. 10 (1966).

\bibitem{Hvelp68}  
 P.~Hvelplund and B.~Fastrup,
 Phys. Rev. \textbf{165}, 408 (1968).

\bibitem{Lennard86}  
 W.N.~Lennard, H.~Geissel, D.P.~Jackson, and D.~Phillips,
 Nucl. Instrum. Methods Phys. Res. B \textbf{13}, 127 (1986).

\bibitem{BrownMoak72}  
 M.D.~Brown and C.D.~Moak,
 Phys. Rev. B \textbf{6}, 90 (1972).

\bibitem{Krist84}  
 Th.~Krist, P.~Mertens and J.P.~Biersack,
 Nucl. Instrum. Methods Phys. Res. B \textbf{2}, 177 (1984).

\bibitem{Garnir80}  
 F.S.~Garnir-Monjoie and H.P.~Garnir,
 J. Phys. France, \textbf{41}, 31 (1980).

\bibitem{LSS63}  
 J.~Lindhard, M.~Scharff, and Schi{\o}tt,
 K. Dan. Vidensk. Selsk. Mat. Fys. Medd. \textbf{33}, no. 14 (1963).

\bibitem{Dib15}  
 A.~Dib, H.~Ammia, M.~Hedibel, A.~Guesmia, S.~Mammeri, M.~Msimanga, and C.A.~Pineda-Vargas,
 Nucl. Instrum. Methods Phys. Res. B \textbf{362}, 172 (2015).

\bibitem{Ziegler77}  
 J.F.~Ziegler,
 Appl. Phys. Lett. \textbf{31}, 544 (1977).

\bibitem{Sigmund04}  
 P.~ Sigmund,
 in \textit{Stopping of Heavy Ions} (Springer Tracts in Modern Physics, 2004), p.~85.

\bibitem{Wilson77}  
  W.D.~Wilson, L.G.~Haggrnark, and J.P.~Biersack,
  Phys. Rev. B \textbf{15}, 2458 (1977).

\bibitem{Pape78}  
 H.~Pape, H.-G.~Clerc, and K.-H.~Schmidt,
 Z. Physik A \textbf{286}, 159 (1978).

\bibitem{AnthLanf82}  
 J.M.~Anthony and W.A.~Lanford,
 Phys. Rev. A \textbf{25}, 1868 (1982).

\bibitem{SchulzBrandt82} 
 F.~Schulz and W.~Brandt,
 Phys. Rev. B \textbf{26}, 4864 (1982).

\bibitem{Abdess92}  
 M.~Abdesselam, J.P.~Stoquert, G.~Guillaume, M.~Hage-Ali, and P.~Siffert,
 Nucl. Instrum. Methods Phys. Res. B \textbf{72}, 7 (1992); \textbf{72}, 293 (1992).

\bibitem{Saga2015}  
 R.N.~Sagaidak, V.K.~Utyonkov, and S.N.~Dmitriev,
 Nucl. Instrum. Methods Phys. Res. B \textbf{365}, 447 (2015).

\bibitem{Trzaska18}  
 W.H.~Trzaska, G.N.~Knyazheva, J.~Perkowski, J.~Andrzejewski, S.V.~Khlebnikova, E.M.~Kozulin, T.~Malkiewicz, M.~Mutterer, and E.O.~Savelieva,
 Nucl. Instrum. Methods Phys. Res. B \textbf{418}, 1 (2018).

\bibitem{Sharma99}  
 Annu Sharma, Shyam Kumar, S.K.~Sharma, N.~Nath, V.~Harikumar, A.P~Pathak, L.N.S.~Prakash Goteti, S.K. Hui, and D.K.~Avasthi,
 J. Phys. G Nucl. Part. Phys. \textbf{25}, 135 (1999).

\bibitem{Perkowski06}  
 J.~ Perkowski, J.~Andrzejewski, A.~Climent-Font, G.~Knya\-zheva, V.~Lyapin, T.~Malkiewicz, A.~Munoz-Martin, and W.H.~Trzaska,
 Nucl. Instrum. Methods Phys. Res. B \textbf{249}, 55 (2006).

\bibitem{ShyKum96}  
 Shyam~Kumar, S.K.~Sharma, N.~Nath, V.~Harikumar, A.P.~Pathak, D.~Kabiraj, and D.K.~Avasthi,
 Radiat. Eff. Def. Solid. \textbf{139}, 197 (1996).

\bibitem{Harikumar96}  
 V.~Harikumar, A.P.~Pathak, S.K.~Sharma, Shyam Kumar, N.~Nath, D.~Kabiraj, and D.K.~Avasthi,
 Nucl. Instrum. Methods Phys. Res. B \textbf{108}, 223 (1996).

\bibitem{ShuKal91}  
 V.~Sch\"{u}le and S.~Kalbitzer,
 Z. Physik A \textbf{340}, 219 (1991).

\bibitem{Harikumar97}  
 V.~Harikumar, A.P.~Pathak, N.~Nath, Shyam Kumar, S.K.~Sharma, S.K.~Hui, and D.K.~Avasthi,
 Nucl. Instrum. Methods Phys. Res. B \textbf{129}, 143 (1997).

\bibitem{LNS68}  
 J.~Lindhard, V.~Nielsen, and M.~Scharff,
 K. Dan. Vidensk. Selsk. Mat. Fys. Medd. \textbf{36}, no. 10 (1968).

\bibitem{Perkowski09}  
 J.~ Perkowski, J.~Andrzejewski, A.~Javanainen, W.H.~Trza\-s\-ka, K.~Sobczak, and A.~Virtanen,
 Vacuum \textbf{83}, S73 (2009).

\bibitem{Echler17}  
 A.~Echler, P.~Egelhof, P.~Grabitz, H.~Kettunen, S.~Kraft-Bermuth, M.~Laitinen, K.~M\"{u}ller, M.~Rossi, W.H.~Trzaska, and A.~Virtanen,
 Nucl. Instrum. Methods Phys. Res. B \textbf{391}, 38 (2017).

\bibitem{Jokinen97}  
 J.~Jokinen,
 Nucl. Instrum. Methods Phys. Res. B \textbf{124}, 447 (1997).

\bibitem{Echler12}   
 A.~Echler, A.~Bleile, P.~Egelhof, S.~Ilieva, S.~Kraft-Bermuth, J.P.~Meier, and M.~Mutterer,
 J. Low Temp. Phys. \textbf{167}, 949 (2012).

\bibitem{SRIMbook}  
 J.F.~Ziegler, J.P.~ Biersack, and M.D.~Ziegler, \textit{SRIM: The Stopping and Range of Ions in Matter}, SRIM Co.,
 Chester, MD (USA), 2008.

\bibitem{Ziegler80}  
 J.F.~Ziegler,
 Nucl. Instrum. Methods, \textbf{168}, 17 (1980).

\bibitem{Barb2010_2}  
 M.~Barbui, D.~Fabris, M.~Lunardon, S.~Moretto, G.~Nebbia, S.~Pesente, G.~Viesti, K.~Hagel, J.B.~Natowitz, and R.~Wada,
 Nucl. Instrum. Methods Phys. Res. B \textbf{268}, 2377 (2010).
 
 \bibitem{Powers68}  
  D.~ Powers, W.K.~Chu, and P.D.~ Bourland,
  Phys. Rev. \textbf{165}, 376 (1968).
  
 \bibitem{Grande88}  
  P.L.~Grande, P.F.P.~Fichtner, M.~Behar, and F.C.~ Zawislak,
  Nucl. Instrum. Methods Phys. Res. B \textbf{33}, 122 (1988).

\bibitem{SagaRange19}  
 R.N.~Sagaidak, N.A.~Kondratiev, L.~Corradi, E.~Fioretto, G.~Montagnoli, F.~Scarlassara, and A.M.~Stefanini,
 Phys. Rev. C \textbf{99}, 014602 (2019).

\end{thebibliography}
\end{document}